\documentclass[%
 rsi,
 amsmath,amssymb,
 reprint,%
]{revtex4-1}

\usepackage{graphicx}
\usepackage{dcolumn}
\usepackage{bm}

\usepackage[utf8]{inputenc}
\usepackage[T1]{fontenc}
\usepackage{mathptmx}
\usepackage{etoolbox}
\usepackage{comment}
\usepackage[dvipsnames]{xcolor}
\usepackage{ amssymb }
\usepackage[caption=false]{subfig}
\usepackage[title]{appendix}
\usepackage{enumerate}
\makeatletter
\def\@email#1#2{%
 \endgroup
 \patchcmd{\titleblock@produce}
  {\frontmatter@RRAPformat}
  {\frontmatter@RRAPformat{\produce@RRAP{*#1\href{mailto:#2}{#2}}}\frontmatter@RRAPformat}
  {}{}
}%
\makeatother
\begin{document}
\preprint{AIP/123-QED}

\title{The QICK (Quantum Instrumentation Control Kit): \\ Readout and control for qubits and detectors}


\author{Leandro Stefanazzi}
 \affiliation{Fermi National Accelerator Laboratory, Batavia IL, United States}
\author{Kenneth Treptow}
 \affiliation{Fermi National Accelerator Laboratory, Batavia IL, United States}
\author{Neal Wilcer}
 \affiliation{Fermi National Accelerator Laboratory, Batavia IL, United States}
\author{Chris Stoughton}
 \affiliation{Fermi National Accelerator Laboratory, Batavia IL, United States}
 \author{Collin Bradford}
 \affiliation{Fermi National Accelerator Laboratory, Batavia IL, United States}
\author{Sho Uemura}
 \affiliation{Fermi National Accelerator Laboratory, Batavia IL, United States}
\author{Silvia Zorzetti}
 \affiliation{Fermi National Accelerator Laboratory, Batavia IL, United States}
\author{Sara Sussman}
 \affiliation{Department of Physics and Department of Electrical Engineering, Princeton University, Princeton NJ, United States}
\author{Salvatore Montella}
 \affiliation{Seconda Università degli Studi di Napoli}
\author{Shefali Saxena}
 \affiliation{GE Healthcare Institute}
\author{Horacio Arnaldi}
 \affiliation{Instituto Balseiro, CNEA, Argentina}
\author{Ankur Agrawal}
 \affiliation{Department of Physics and Pritzker School of Molecular Engineering, University of Chicago, Chicago IL, United States}
\author{Helin Zhang}
 \affiliation{Department of Physics and Pritzker School of Molecular Engineering, University of Chicago, Chicago IL, United States}
\author{Chunyang Ding}
 \affiliation{Department of Physics and Pritzker School of Molecular Engineering, University of Chicago, Chicago IL, United States}
\author{Andrew Houck}
 \affiliation{Department of Physics and Department of Electrical Engineering, Princeton University, Princeton NJ, United States}
\author{David I Schuster}
 \affiliation{Department of Physics and Pritzker School of Molecular Engineering, University of Chicago, Chicago IL, United States}
\author{Gustavo Cancelo}
 \affiliation{Fermi National Accelerator Laboratory, Batavia IL, United States}

 \email{cancelo@fnal.gov.}

\date{\today}

\begin{abstract}
We introduce a Xilinx RFSoC-based qubit controller (called the Quantum Instrumentation Control Kit, or QICK for short) which supports the direct synthesis of control pulses with carrier frequencies of up to 6 GHz. The QICK can control multiple qubits or other quantum devices. The QICK consists of a digital board hosting an RFSoC (RF System-on-Chip) FPGA, custom firmware and software and an optional companion custom-designed analog front-end board.  We characterize the analog performance of the system, as well as its digital latency, important for quantum error correction and feedback protocols. We benchmark the controller by performing standard characterizations of a transmon qubit. We achieve an average gate fidelity of $\mathcal{F}_{avg}=99.93\%$. All of the schematics, firmware, and software are open-source.
\end{abstract}

\maketitle

\section{Introduction}\label{Intro}

\begin{figure*}[!ht]
    \includegraphics[width=0.99\textwidth]{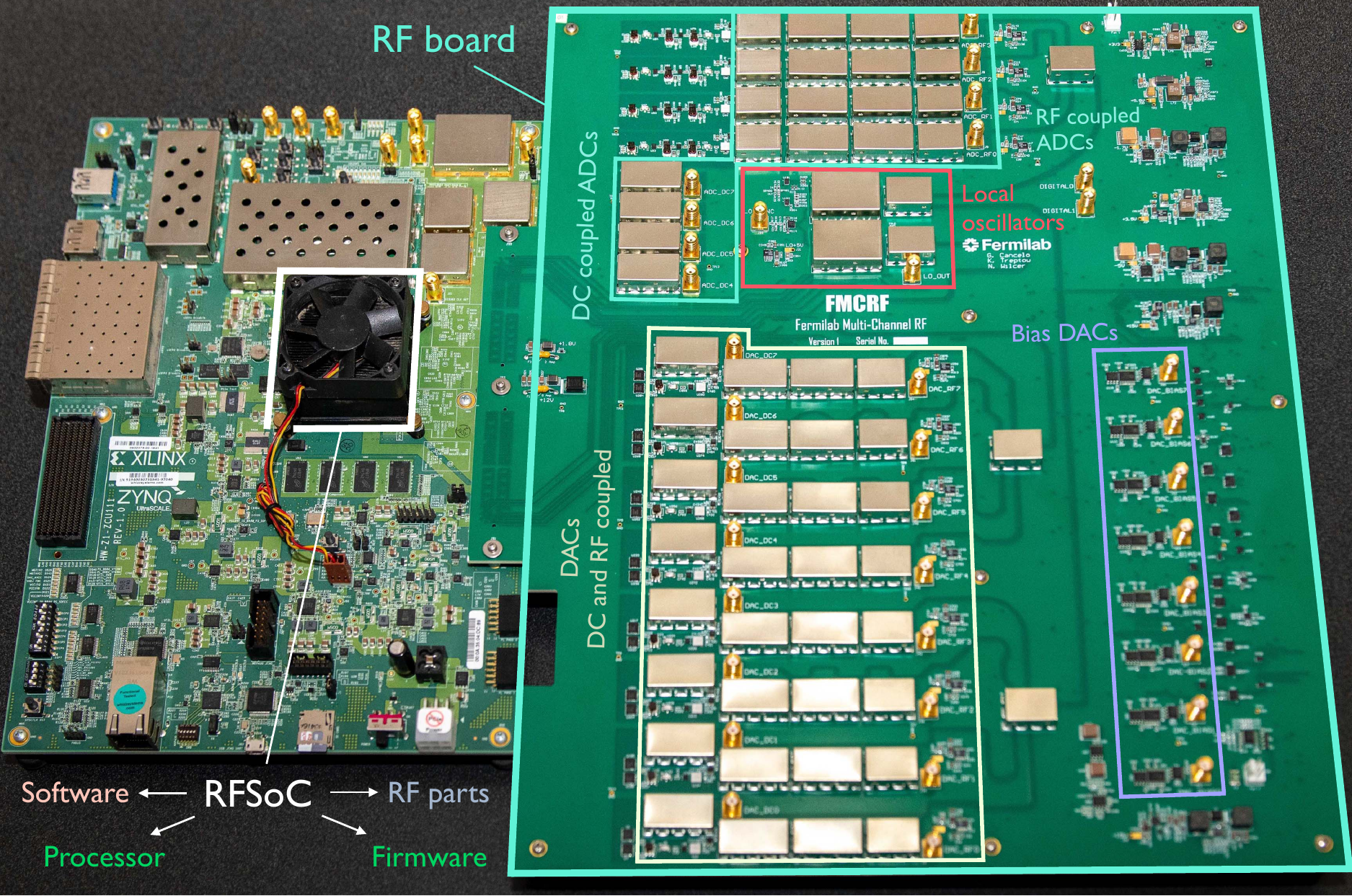}    
    \caption{The Quantum Instrumentation Control Kit (QICK). The QICK consists of two pieces of hardware: the commercial ZCU111 RFSoC evaluation board (left), which connects to the QICK RF board (right) which can be used for additional signal up/downconversion, amplification and filtering.}
    \label{fig:rfsocboard}
\end{figure*}

Quantum computers are predicted to outperform classical computers in problem domains such as decryption \cite{Shor_1997, Gidney2021}, secure communication \cite{Bennett_2014, Yin2020}, quantum chemistry \cite{Wang_2020} and machine learning \cite{Biamonte_2017}. Quantum bits (qubits), have been developed in several different platforms including trapped ions \cite{Egan_2021}, superconducting qubits \cite{Kjaergaard_2020}, semiconductor quantum dots \cite{Yoneda_2017}, color centers \cite{Schirhagl_2014}, and neutral atoms \cite{Bloch_2005}.  In all of these systems, the ability to synthesize a large number of control signals, measure the states of the qubits, and perform feedback in real-time is a critical requirement. Historically, this has been done using either expensive general-purpose test equipment or proprietary embedded systems.  This has made controlling even moderate size quantum processors with tens of qubits, such as those hosted by IBM or Google \cite{Sager_2020, Arute2019}, prohibitive for academic labs and small startups.  We have developed a RF/FPGA-based system that is flexible enough to support different platforms and is affordable to academic laboratories.

There have been several platforms developed which integrate fast RF DACs and ADCs with Field-Programmable Gate Array (FPGA)s.  Early academic efforts in the superconducting qubit community were made at Delft for pulse routing \cite{Asaad_2016} and at ETH Zurich for real-time processing of measurements \cite{Walter_2017}. The first demonstration of quantum error correction to show improvement over the physical constituents was enabled by an FPGA controlled data acquisition system \cite{Ofek2016}. The ion trap community has developed the FPGA-based ARTIQ system \cite{artiq} for control of their systems, albeit on slower timescales than those required for solid-state qubits, which typically have faster gate speeds (and decoherence rates).  Commercial products with FPGA-enabled real-time pulse synthesis and readout have recently become available from vendors like BBN \cite{Ryan2017}, Keysight \cite{keysight}, Zurich Instruments \cite{zurichinst}, and Quantum Machines \cite{quantummachines}. An open-source qubit controller was recently developed by LBNL and UC Berkeley \cite{xu2021}. These solutions all use a conventional method by which the FPGA-controlled DAC synthesizes the IF envelope and is upconverted via IQ mixing with a local oscillator supplied by an analog RF source.  

More recently, a newer generation of RF DACs operates at high enough sampling rates that it becomes possible to directly synthesize microwave pulses without any kind of upconversion \cite{Kalfus2021}, eliminating the need for meticulous calibration \cite{Jolin2020}. In this work we take advantage of a newly developed FPGA platform, the RFSoC from Xilinx \cite{rfsocwebsite}, which integrates high-speed DACs, ADCs, programmable FPGA logic, and a conventional microprocessor in the same package. RFSoC-based qubit controllers have been homemade by several academic labs \cite{Gebauer2021} \cite{park2021icarusq} and one was made available commercially by Intermodulation Products \cite{imp}. 

The Quantum Instrumentation Control Kit (QICK) is an open-source RFSOC-based platform which combines FPGA firmware for real-time processing and a Python interface running the PYNQ \cite{pynqwebsite} operating system.  The initial implementation of the QICK uses the ZCU111 \cite{rfsocdatasheet} evaluation board with eight DACs and ADCs.  This allows pulse synthesis up to 6 GHz directly using the XM500 RFMC balun card provided in the evaluation kit, and higher by using either external sources and mixers or the custom RF board we have developed. The system can synthesize and digitally upconvert arbitrary pulses, measure and digitally downconvert incoming signals, and perform real-time decisions and feedback based on the input.  The system is open-source \cite{QICKrepo} and can run on commercially available hardware at a price significantly cheaper than commercial offerings at this time.  

The rest of the manuscript is organized as follows: in section \ref{sec:hardware} we describe the hardware and its capabilities.  In section \ref{Sysarchitecture} we describe the controller architecture, and the micro-sequencer with precision timing blocks, along with the digital upconversion and downconversion chains.  Next, in section \ref{RFpnm} we characterize the analog performance of the direct output of the ZCU111 using both the XM500 RFMC balun card and the custom analog QICK RF board. We conclude section \ref{RFpnm} by demonstrating the use of the system to fully characterize a superconducting qubit previously used to perform a qubit-enhanced search for dark matter \cite{Dixit2021}. The custom RF board has not yet been used in qubit experiments due to part shortages and supply chain delays. Instead, discrete RF hardware was used as shown in section \ref{sec:qubit} Figure \ref{fig:wiring}.

\section{Digital and Analog Hardware}\label{sec:hardware}

The QICK (Quantum Instrumentation Control Kit) is a multi-input, multi-output high performance controller designed for qubit systems and superconducting detectors. The QICK can be used as a flexible instrument to control and characterize new qubits and detectors or it can be used as a module in a multi module architecture for a large detector instrument or quantum computer. A precursor to the QICK is the FNAL-Gen2 fMESSI system which generated two chronograph instruments: DARKNESS and MEC. MEC uses a stack-up of 20 fMESSI boards and it has been operating at the 8 meter Subaru telescope since 2018 \cite{Walter2020}. The QICK is designed to be self-contained: the user directly connects RF lines between the QICK and the dilution refrigerator containing the qubits.  With appropriate firmware the QICK hardware can also be used for superconducting detectors such as MKIDs (Microwave Kinetic Inductance Detectors) or TESes (Transition Edge Sensors) connected to RF microchips for high-density frequency multiplexing. Since the RFSoC analog inputs each have 2 GHz of continuous usable bandwidth, the number of detectors per channel depends on the channel bandwidth and separation. A typical application \cite{Fruitwala2020} has 1000 MKID channels separated by 2 MHz. The QICK could be used to read out several thousands of multiplexed detectors, a number that will ultimately be limited by the available FPGA logic in the specific design. MKIDs and TESes coupled to microwave cavities require an excitation tone at the resonance frequency of the pixel, and so the QICK DACs will be used to provide the same number of tone excitations. Using the QICK as a MKID/TES controller will be described in a separate paper. 

The QICK system is simpler, more compact, and cheaper than the conventional qubit control stack that has many auxiliary components (e.g. two DACs and one external local oscillator per channel).  The QICK hardware consists of two parts, a main evaluation board and an optional custom-designed analog front-end shown together in Fig.~\ref{fig:rfsocboard}.  

\subsection{Xilinx ZCU111 RFSoC evaluation board}\label{RFSOC_FPGA}

\begin{figure}[h!]
\begin{center}
\includegraphics[width= 1\columnwidth]{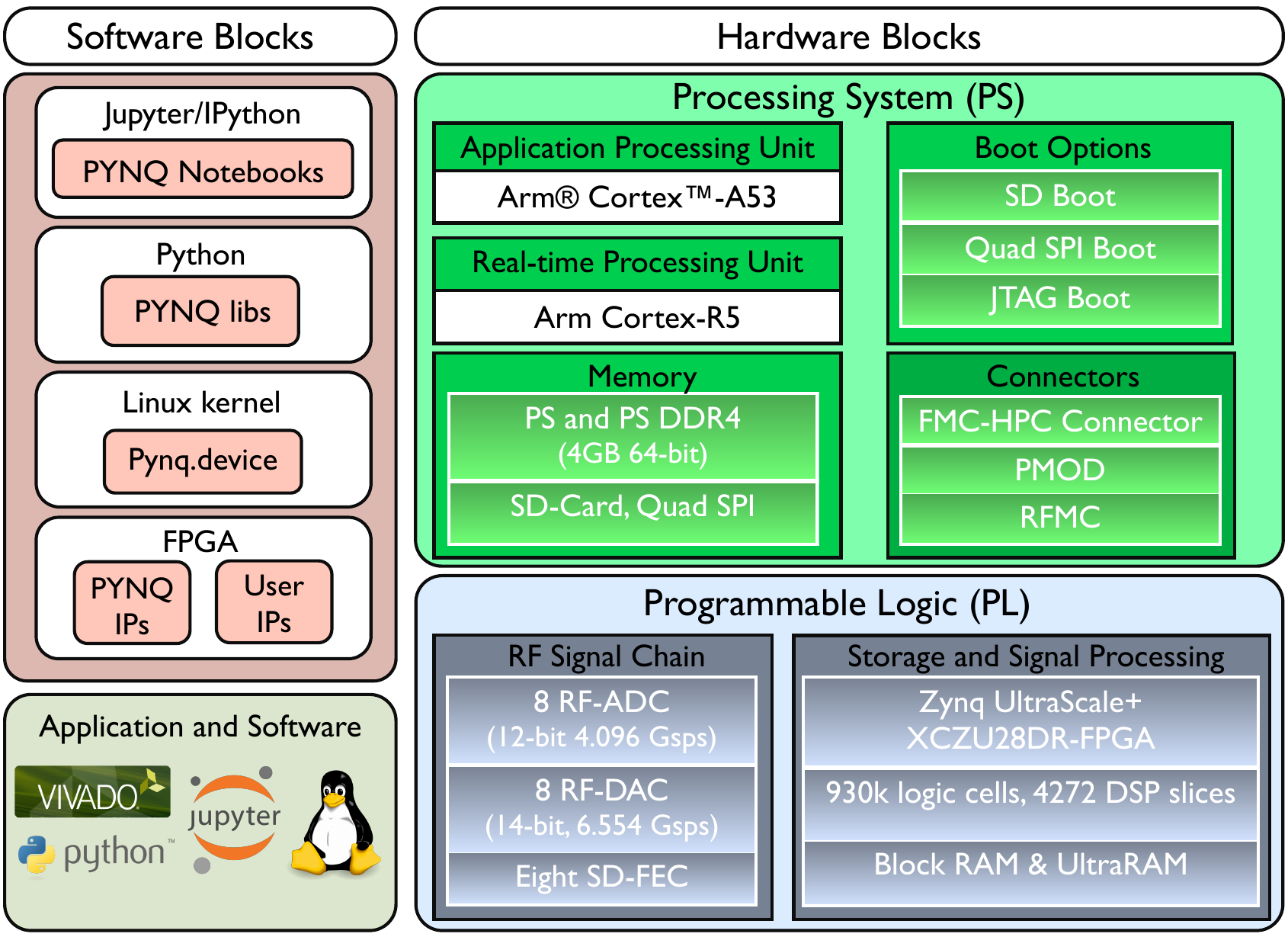}
        \caption{The Xilinx XCZU28DR RFSoC chip block diagram. The chip consists of software blocks and hardware blocks. The software blocks can be subdivided into lower-level blocks such as the Linux kernel, and higher-level blocks such as the PYNQ software library whose functions are called from Jupyter notebooks. The hardware blocks can be subdivided into blocks related to the processing system (PS) and the programmable logic (PL).}
        \label{fig:rfsocbd}
\end{center}
\end{figure}

The QICK takes advantage of the highly integrated RFSoC FPGA. The XCZU28DR RFSoC chip (Fig.~\ref{fig:rfsocbd}) has eight 6.5 GS/s digital to analog converters (DACs) and eight 4 GS/s analog to digital converters (ADCs). Both the DAC and ADC blocks include configurable IQ digital up/down conversion, an integrated numerically controlled oscillator (NCO), a gain matrix, and digital filters with interpolation/decimation. They are easily integrated to the logic through standard AXI interfaces and avoid the use of high-power drivers needed with external A/D and D/A devices that require LVDS or JDEM interfaces.The RFSoC also contains several different ARM processors, 70 Mb of memory and multiple interfaces. 

Most commercial qubit controllers have DACs with <1 GHz of analog bandwidth, so RF qubit control pulses (typically 4-6 GHz) must be upconverted with analog mixers. In contrast, the qubit controller presented here can directly synthesize carrier frequencies of up to 3 GHz in first Nyquist zone mode and it can directly synthesize carrier frequencies of up to 6 GHz in second Nyquist zone mode. This eliminates mixer spurs driving undesirable transitions, and eliminates the need to carefully calibrate IQ mixer offsets and gains.

\begin{figure}[!hb]
    \centering
        \includegraphics[width=0.9\columnwidth]{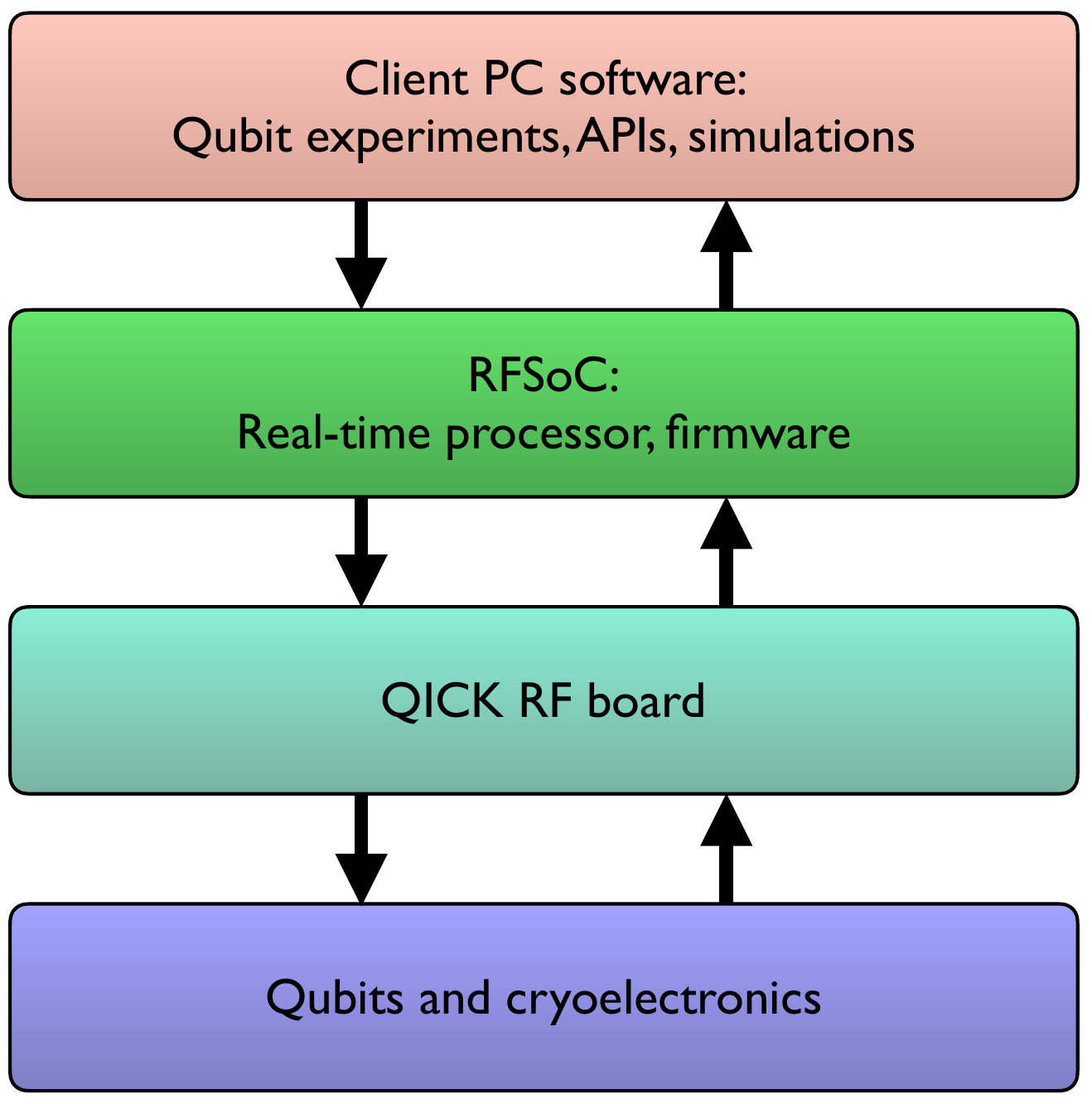}
    \caption{The QICK stack-up. The user runs qubit experiments via the QICK Python API (top level of the stack). The qubit experiment is sent to the RFSoC (second level of the stack) and translated into FPGA-level instructions. Signals generated by the RFSOC are processed further by the QICK RF board (third level of the stack) and then sent to to the qubits (the bottom level of the stack). Qubit measurements are then read into the QICK in the reverse order of the stack-up and the measurement results are returned by the Python API. Up to an entire quantum program can be executed by the RFSoC without high level software calls.}
    \label{fig:stackup}
\end{figure}

\subsection{Analog Front-end} \label{AC_DC}

\begin{figure}[!ht]
    \centering
        \includegraphics[width=1\columnwidth]{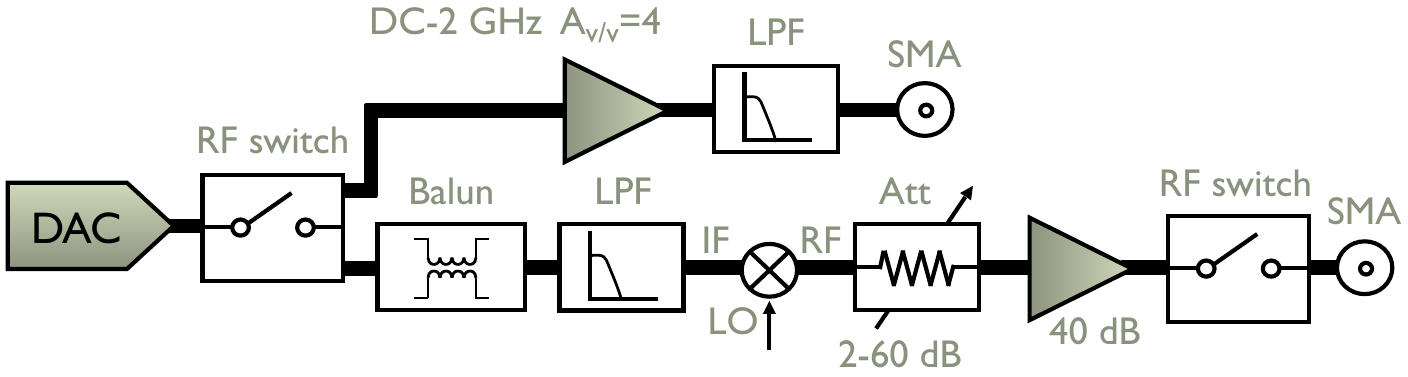}
    \caption{The QICK RF board DAC output schematic. In software, the user can select the DAC output to be RF or DC coupled. The output signal is then processed by the associated DAC output chain. Note that the two step attenuators and three RF amplifiers are interleaved to optimize power linearity.}
    \label{fig:DAC}
\end{figure}

\begin{figure*}[!]
    \centering
    \includegraphics[width= 0.99\textwidth]{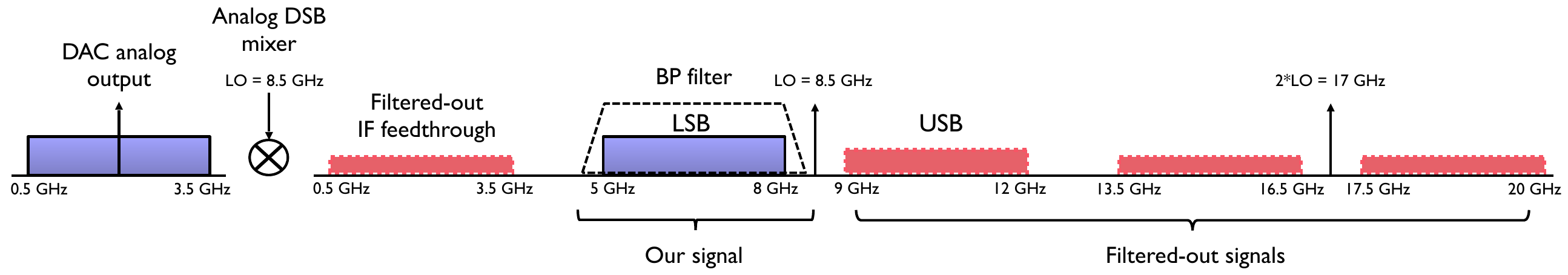}%
    \caption{The QICK RF board upconversion schematic. In this example, the QICK RF board LO is parked at 8.5 GHz. The user can place the IF signal anywhere in the 0.5-3.5 GHz band and then it is upconverted into the 5-8 GHz band. The QICK RF board filters the spurs in the unwanted sidebands. Although the fast digital DDS can generate a carrier lower than 0.5 GHz, a minimum of few hundred MHz is desired to avoid getting too close to the edge of the filter that suppresses the LO and higher frequency signals. The current version of the RF output includes a Mini-Circuits FCN-1800+ low pass filter that can be replaced if more analog bandwidth is needed.}
    \label{fig:upconversion}
\end{figure*}


The QICK RF board contains more than 200 components, including amplifiers, mixers, filters, local oscillator generators, switches and drivers. All RF and DC coupled outputs/inputs are accessible via SMA connectors. The QICK only requires a 50 watt, 12 volt DC power supply. In the following sections we describe the integrated RF/FPGA chip at the center of the controller, the RF board's RF and DC coupled outputs and inputs, and the RF board's components which are used for bias and digital I/O.

The QICK has an integrated low jitter master clock; however, a stack-up of multiple QICK boards can be synchronized to a single external stable reference. See Appendix \ref{Two board synchronization} for details of clock synchronization measurements. Fig.~\ref{fig:stackup} shows the high-level block diagram of the controller which will be described in detail in Section \ref{Sysarchitecture}.

The custom RF board, (larger board on the right side of Figure \ref{fig:rfsocboard}) extends the eight RFSoC DAC outputs to either RF or DC coupled amplification and filtering. A simplified block diagram of the output chain is shown on Fig.~\ref{fig:DAC}. Every DAC output is connected to a software-controllable switch that allows the user to choose between an RF output or a DC coupled output. The RF output path upconverts the RFSoC DAC output signal using the MM1-0212HSM double balanced mixer, which operates between 2-12 GHz. The RF board's amplifying chain has been optimized to have the highest performance between 3-8 GHz (Fig.~\ref{fig:upconversion}). The output of the mixer (typically $\sim-23$ dBm) is amplified by 40 dB and attenuated by two digital step attenuators. The step attenuators introduce a minimum insertion loss of 1 dB and total a maximum attenuation of up to 60 dB in 0.25 dB steps. So, the RF output power dynamic range is 4 to $-56$ dBm. Alternatively, the DAC output can be  switched to a DC coupled amplifier with 2 GHz of bandwidth. The main purpose of the DC coupled output is to control fast unmodulated signals such as fast flux pulses for fluxonium qubits as in \cite{Zhang2021} or for other quantum systems which integrate fast voltage or flux pulses, such as spin qubits in quantum dots \cite{Hanson2007}.

\begin{figure}[h!]
    \centering
    \subfloat[]{{\includegraphics[width=1\columnwidth]{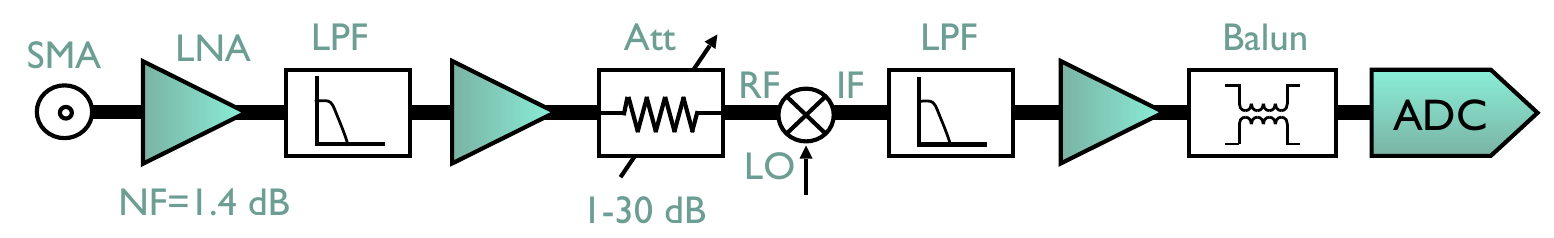} }}%
    \\
    \subfloat[]{{\includegraphics[width=0.5\columnwidth]{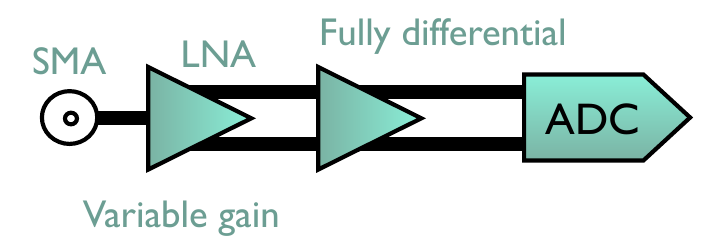} }}%
    \caption{The QICK RF board ADC input schematic. The QICK has four dedicated RF inputs (a) optimized for 4-8 GHz, and four dedicated DC coupled inputs (b) with a bandwidth of 1.5 GHz.}
    \label{fig:ADC}
\end{figure}

Four of the eight QICK RF inputs (Fig.~\ref{fig:ADC}) are designed for RF signals, and the other four are DC coupled with an analog bandwidth of 1.5 GHz (for use as auxiliary oscilloscope or spectrum analyzer inputs, for example). The four RF inputs are designed to amplify low noise RF signals coming from the dilution refrigerator. The noise temperature of the RF input channel is governed by the noise figure of the first amplifier in the chain, the MACOM MAAL-011130, which is 1.4 dB. The input signal is amplified by four high-gain, high-P1dB compression amplifiers and attenuated by a step attenuator with a maximum of 30 dB in 0.25 dB steps. The RF input chain is mixed down using the same Marki mixer used for upconversion. So, the RF input dynamic range is $-60$ to $-90$ dBm. Note that the 4 GHz analog bandwidth of the RF ADCs suggests that we can frequency multiplex the readout of several qubits onto the same line. If we were to have 100 MHz bandwidth readout cavities that were separated by 200 MHz, 10 qubits could be simultaneously read out by the same RF ADC with room to spare.

The QICK RF board also includes additional DACs and digital I/O that are separate from the RFSoC. There are eight 20-bit DAC outputs for the biasing and DC control of flux for quantum devices that are tuned by flux or voltage such as flux qubits or spin qubits in quantum dots. The bias maximum output voltage is $\pm$ 10 volts with 1 ppm resolution, 1 ppm integral nonlinearity, 7.8 nV /$\sqrt{\text{Hz}}$ of white Gaussian noise and a 1/f noise knee at 4 Hz. There are also 16 digital software-configurable bidirectional I/O for the purpose of synchronization and triggering between the QICK and other instrumentation.


The on-board local oscillators (LOs) for the 12 RF mixers (eight upconverters and four downconverters) are generated by three LMX2595 PLL and fractional frequency synthesizers. Each PLL has two outputs which are split in two by a Mini-Circuits EP2C+ splitter. Each PLL can be set to a different LO, allowing for two sets of four DACs and one set of four ADCs to have separate LO frequencies. Each on-board LO can be tuned from DC to 15 GHz.

The PLL output is amplified by an HMC788 to achieve optimum LO power at the mixers for a broadband range of $\sim$ 1 GHz to 10 GHz or more. The high frequency (10 kHz-10 MHz) integrated jitter is 55 fs. The 10 Hz-10 MHz integrated jitter is $\sim$ 500 fs. 

Appendix \ref{bandwidths} shows a Table with output and input bandwidths of the QICK system.

\section{System architecture and functionality}
\label{Sysarchitecture}

\begin{figure}
\begin{center}
\includegraphics[width=1\columnwidth]{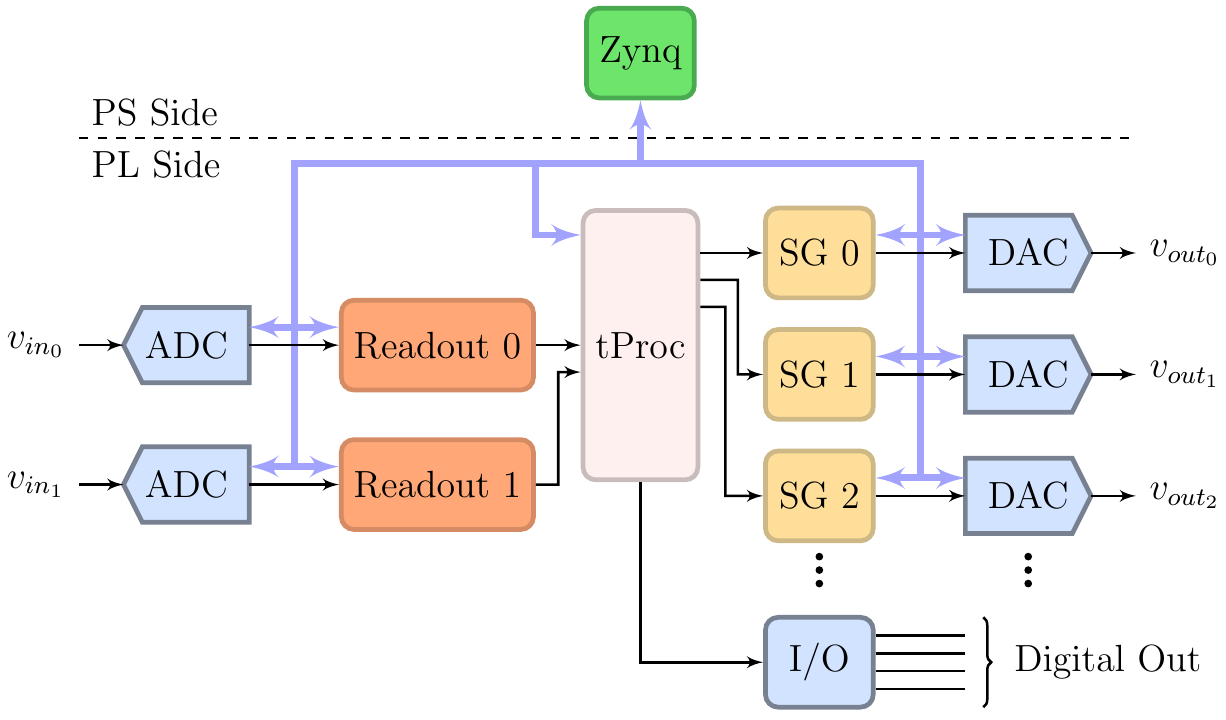}
        \caption{The QICK firmware block diagram. The QICK firmware mainly consists of the timed-processor (tProcessor) block (Figure \ref{fig:tProc}), the signal generator (SG) blocks (Figure \ref{fig:SG}) and the readout blocks (Figure \ref{fig:Readout}). Instructions are passed between the RFSoC Zynq processor and the firmware blocks. These instructions cause signals to be sent and received from the RF blocks (DAC blocks, ADC blocks, and digital output blocks).}
        \label{fig:fbd}
\end{center}
\end{figure}

The functionality of the QICK system is divided between the Processing System (PS) and Programmable Logic (PL), shown in Fig.~\ref{fig:fbd}.  The PS of this UltraScale+ device is a ZYNQ system with its own DDR4 memory, which runs the Linux operating system on a multi-core ARM processor. The PS uses PYNQ libraries and drivers to enable direct memory access (DMA) to the PL. The user interface is typically via Jupyter notebooks accessed from a remote web browser. Firmware in the PL includes signal generator blocks and readout blocks controlled by a timed-processor (tProcessor) block to implement time-critical functions.

Data flow between the PS and PL uses AXI interfaces. Fast data transfers are implemented using DMA logic on the PL side. The firmware version has a companion driver file, which exposes all features to the user as a simple collection of Python classes and objects \cite{QICKdoc}. The user can define experiments directly from Python, leaving all low-level hardware details up to the driver interface.

The core of the QICK system is the tProcessor. This block implements a custom processor, with the addition of timed instructions to ensure events are executed at proper times. Pulses for the control and readout of qubits are specified with a complex envelope which modulates a high-speed carrier. This functionality is implemented by the signal generator block. These blocks have internal memories to upload the IQ tables, and fast parallel DDS blocks for digital upconversion. As Section \ref{signalgenerator} explains, the main advantage of the fast DDS is that it generates a tone between 0 and 3 GHz. The $n$th order harmonics of that tone are typically far away from qubit/cavity resonances and can also be easily filtered out rather than calibrated down with an IQ mixer. The tone images from folding around the Nyquist frequency bands are at predictable locations and so the user can frequency-plan to make sure those are far from their signal and thus filtered out. The output of each ADC is connected to a readout block, which implements fast digital downconversion and eight times decimation and filtering, followed by averaging and buffering for signal detection. An additional digital output I/O channel carries digital markers which are routed to physical connectors to allow for external equipment triggering or fast analog switch control.

The QICK system described in this paper has a fixed number of the three main kinds of blocks:  tProcessor, signal generator, and readout.  These components have been designed to allow flexibility to address evolving demands.  This paper describes the initial version of the firmware.  As new configurations are available, these versions can be dynamically loaded between experiments, if necessary, using high-level Python commands.

In a typical experiment, the user will load the waveforms into the signal generator blocks from the PS to the PL. Each readout block is configured before launching the experiment. The program for the tProcessor is uploaded before it begins to run. Even when the tProcessor is programmed with a low-level custom defined assembly language, quantum programs are written using high-level Python classes to ease the description of the experiment. The QICK Github repository and QICK documentation website provide demonstrations and documentation of these classes \cite{QICKrepo,QICKdoc}. This allows users to directly access all of the capabilities of the system in a user-friendly environment. An effort to make QICK a backend compatible with standard quantum programming languages such as Qiskit is underway \cite{qiskit}.

\subsection{Timed-processor}
\label{programcontrol}
\begin{figure}
    \centering
    \includegraphics[width=1\columnwidth]{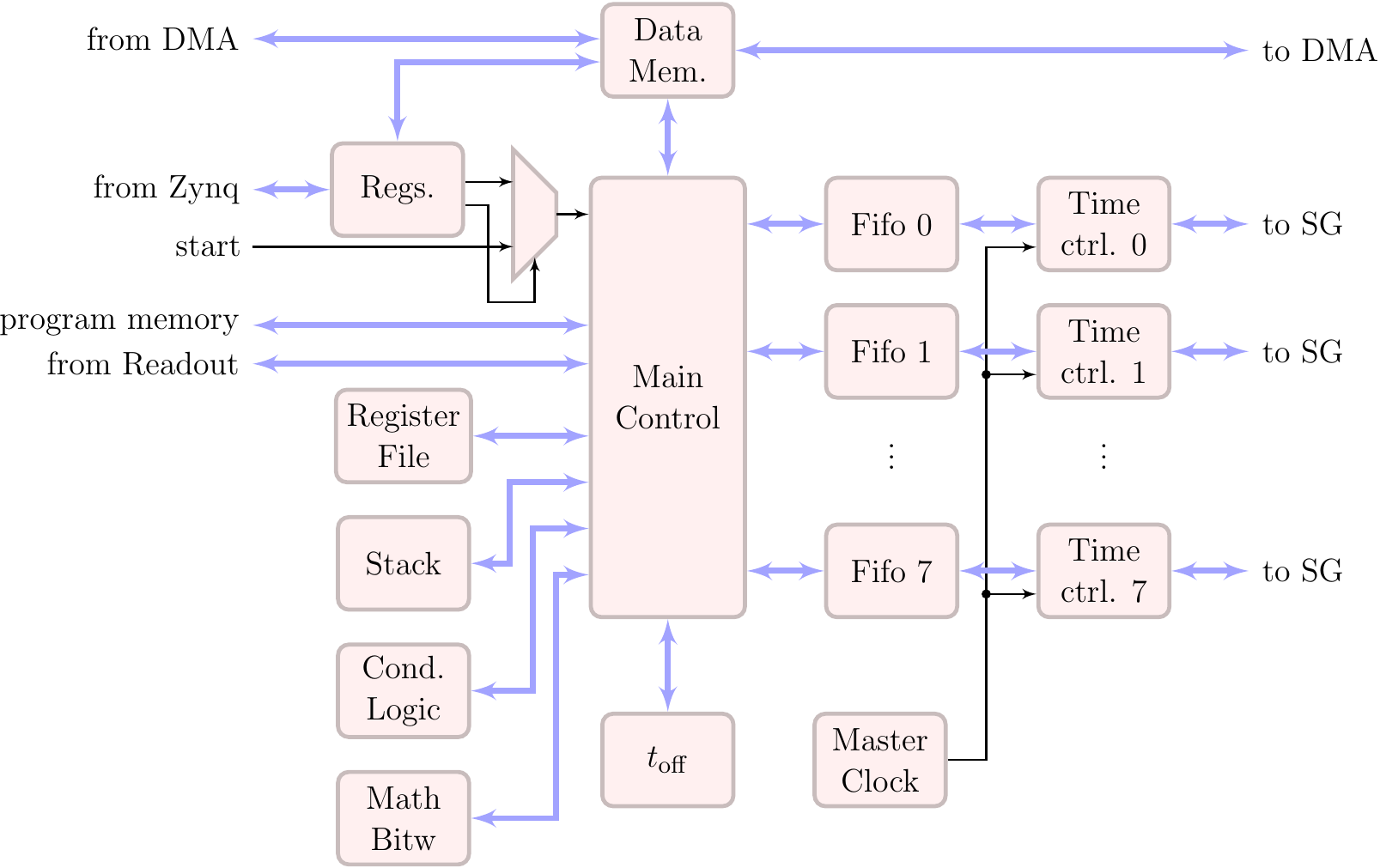}
    \caption{The microsequencer (tProcessor) firmware block diagram. The tProcessor has a master clock and a time offset register $t_\text{off}$ that is used to align user instructions in time. Timed instructions are dispatched to the signal generator (SG) (Figure \ref{fig:SG}) and readout blocks (Figure \ref{fig:Readout}). Also, the user can execute common operations like register read/write, addition and subtraction, bitwise number manipulation, loop, and memory access. The tProcessor block has a dedicated data memory which is accessible from the PS using a single AXI read/write or fast DMA transfers. This memory is accessible from the tProcessor itself and can be used as data exchange or for parameters of the experiment.}
    \label{fig:tProc}
\end{figure}

Figure~\ref{fig:tProc} shows the block diagram of the 64-bit timed-processor (tProcessor).
This block is a custom processor with the addition of timed instructions. The tProcessor implements an assembly language grammar including looping, conditional branching, register access, signal generation, and readback. The tProcessor is programmed and controlled with Python APIs. %

The tProcessor uses a single master clock which is implemented using a 48-bit counter. This means that the user can run a single tProcessor program for up to eight days continuously for the implementation presented here. This number could be further increased to allow for longer experiments.

When an instruction with a time tag is found in the program memory, this instruction is dispatched into the corresponding queue of the specified channel (the processor integrates eight output channels). The processor decoding and execution is decoupled from the queue as it can look ahead of time to accelerate the software flow. Instructions dispatched to the timed instruction control queue are executed at the time instructed by the time tag. When the time arrives, the instruction is executed by the timed instruction control sub-block. If the queue of a particular channel gets full, the processor will wait until space becomes available before going further in the program. In other words, non-timed instructions are executed as in a standard processor and timed-instructions are dispatched to a queue, which is handled by the time control logic.  A more detailed description of the timelines relevant to the QICK architecture is given in Appendix \ref{Timelines}. 

To lower the number of bits necessary to encode the time of a certain instruction, the tProcessor has a special register that indicates the time offset of the time axis. This time offset register is manipulated with specific instructions. When the processor starts, the time offset is set to zero. If a timed instruction is found, it is dispatched with its time tag as the absolute time. As the experiment advances, the time offset register is updated by specific instructions. For example, suppose that the time offset is now 100, and that the same timed instruction is found. Then the absolute time of the instruction will be 100 plus its time tag.  This is dispatched into the queue. This simple structure allows the user to describe times in a very simple manner. Output channels are parallelized, so two or more channels could execute distinct instructions at the same time. Multiple tProcessors could be added to the firmware in the future to control more qubits, as they are designed for an external, synchronized start.

\subsection{Signal generator}
\label{signalgenerator}
\begin{figure}
    \centering
    \includegraphics[width=1\columnwidth]{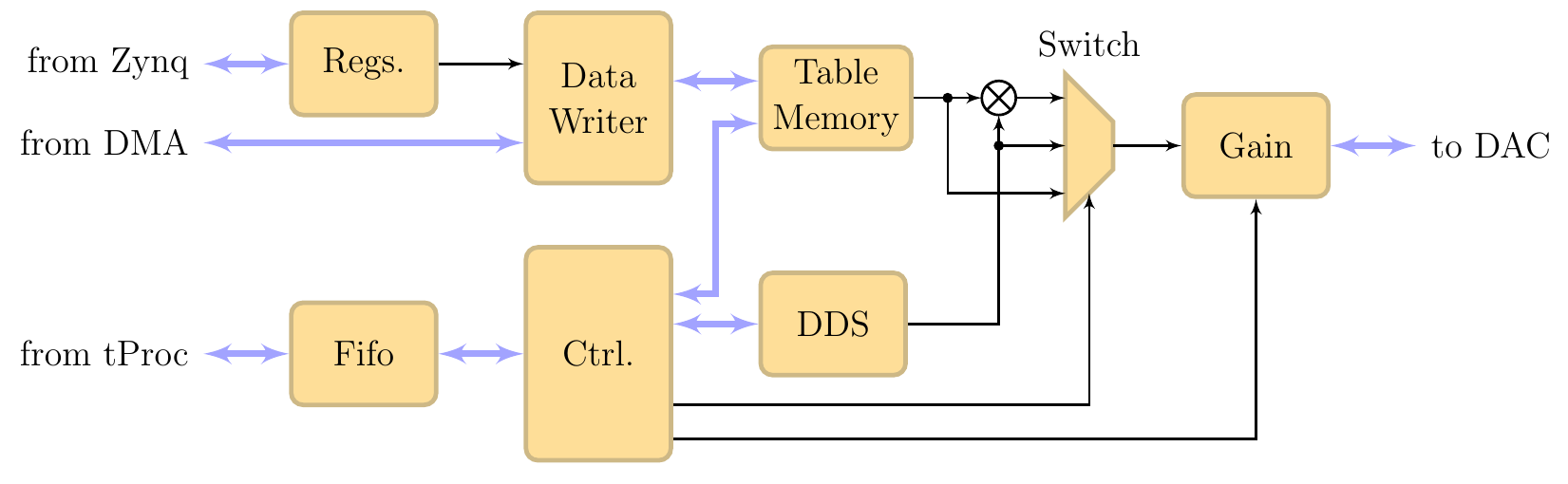} %
    \caption{The signal generator firmware block diagram. The user sends waveform parameters via Python (running on the Zynq PS) to the signal generator block. These waveforms are played at the time specified by the tProcessor block (Figure \ref{fig:tProc}) to ensure phase coherence. The table memory block stores the I and Q values for the waveform envelope. The fast DDS block synthesizes the tone used for digital upconversion. The switch determines which mode of upconversion to do. The signal then passes through a gain block before entering the DAC block.}
    \label{fig:SG}%
\end{figure}

Figure~\ref{fig:SG} shows the signal generator FPGA block diagram. The signal generator plays pulse envelopes from a library of pulses created by the user. This block works as an always-ready slave, meaning that if no pulses are queued to be played, its output is chosen to either be zeroed or kept constant from the last played sample. Pulses are played at the time specified by the tProcessor, which guarantees phase coherence. If the specified time is 0, pulses are played immediately. If the queue has been empty and a new pulse is pushed in, there is a minimum latency of 20 clocks. The tProcessor uses a standard AXI stream interface to ease system connectivity.%

The generator has two main blocks: table memory and DDS. The table memory, which holds I and Q values for the pulse envelope, uses the FPGA's internal BRAM memories. To meet timing for the high-speed DAC, data are interleaved 16-bit words for I and Q. Because the maximum FPGA clock speed is 16 times slower than the maximum DAC speed, the DDS block is highly parallelized to enable the generation of waveforms at the maximum DAC speed. Memory output samples and DDS outputs are complex-multiplied to implement digital upconversion. A single instruction fully specifies the waveform in terms of the start address of the envelope, DDS frequency and phase, pulse duration, output selection and gain. This allows the output to, for example, switch efficiently from a Gaussian pulse of 20~ns duration and carrier frequency 2~GHz to a pulse of 1~$\mu$s and no high-frequency modulation, and then back to a square envelope pulse at carrier frequency $2.8$~GHz. The resolution for the fast DDS is 32-bit, which gives $\sim 1.5$~Hz of resolution with a sampling frequency of 6~GHz on the output DAC. It is worth noting that even when the DAC transfers are done at the FPGA clock rate of 384 MHz and 16 samples at a time, the envelope samples can encode up to the full Nyquist bandwidth of the DAC, which in IQ represents 6 GHz.

An important feature of the signal generator block is phase coherence.  For proper control of qubits, each signal generator associated to a DAC controls the output pulse phase as follows: an output pulse is sent for implementing a particular rotation of the qubit at a frequency $\omega_0$ and then a second pulse is sent with a different frequency $\omega_1$.  If the experiment needs to send another pulse at frequency $\omega_0,$ the phase of that waveform needs to be coherent with the previous $\omega_0$ pulse. An easy way of visualizing this property is thinking of a continuous set of sine wave generators that never stop. Then a phase for a particular frequency at any time corresponds to the phase of the continuously running sine wave of that frequency. In reality, at the time a pulse is output, the signal generator block computes the required phase for a particular frequency with respect to the continuously running sine wave synchronized to master clock origin of time. That allows to have frames with different frequencies on the same signal generator.

The \texttt{mode} of the signal generator can be one-shot or periodic.  If \texttt{mode=0} the signal generator will create a waveform with the specified number of samples one time.  If \texttt{mode=1} the signal generator will keep repeating the actual waveform until the next waveform is read from the queue.  The value of \texttt{stdsel} determines behavior after all samples are generated. If \texttt{stdsel=0} the last value of the pulse repeats, and if \texttt{stdsel=1} the output is set to 0 once all waveforms are completed.

The output of the signal generator is controlled by the switch using the \texttt{outsel} parameter. The user can set \texttt{outsel=0} outputs the complex mixing of the DDS and table memory; \texttt{outsel=1} outputs the DDS; \texttt{outsel=2} outputs the table memory; \texttt{outsel=3} outputs 0.  This result is multiplied by a gain and sent to the DAC.

\subsection{Readout} \label{Readout_Control_architecture}
\begin{figure}
    \centering
    \includegraphics[width=1\columnwidth]{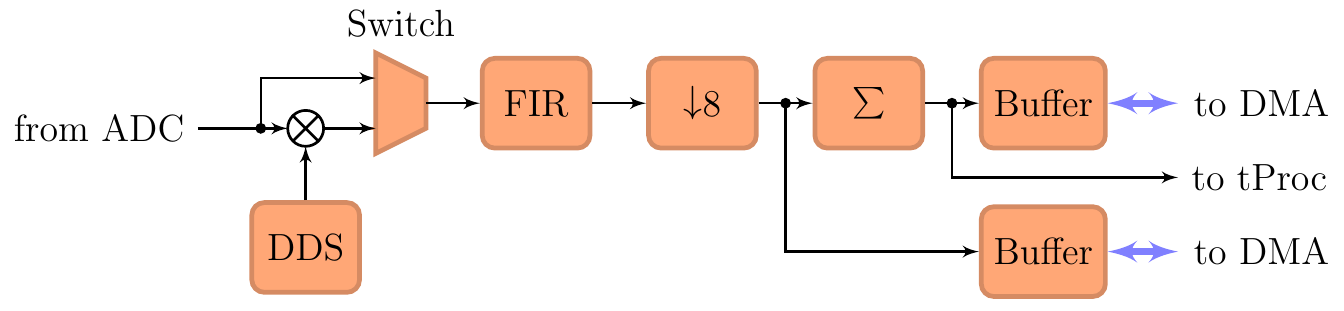} %
    \caption{The readout firmware block diagram. Incoming samples come at high-speed from the ADC block to the readout block. Digital downconversion is performed by multiplying the high-speed incoming samples with a fast DDS. The switch determines which mode of downconversion to do. The resulting signal is low-pass filtered and decimated by 8. The signal is then captured and accumulated in buffers, a process which is controlled by the tProcessor block (Figure \ref{fig:tProc}) readout trigger.}
    \label{fig:Readout}%
\end{figure}

The readout block is shown in Fig.~\ref{fig:Readout}. The input samples are provided by the parallel digital interface of the high-speed ADC. This block provides eight samples in parallel to allow operating at FPGA speeds. 

Similar to the signal generator block, digital downconversion is performed by multiplying the high-speed incoming samples with a fast DDS structure. This parallel DDS has eight individual DDS blocks so that the entire ADC bandwidth is covered in the downconversion process. A switch block allows the user to bypass the downconversion and read the raw data. Upon downconversion the signal is low-pass filtered and decimated. Filtering and decimation may be tailored to a particular experiment. The filtered signal is fed into the average and raw-buffer blocks. The raw-buffer is mostly used for debugging as it captures the raw, decimated samples. Instead, the average block captures an I and Q pair per readout. The average and capture process is started using a trigger controlled by the tProcessor block. Subsequent triggers will repeat the same average and buffer operation. The averaged values are stored in consecutive memory locations of a circular buffer memory. Data from this buffer is read by the PC using fast DMA transfers. Readout triggers can be executed inside a tProcessor loop, allowing for repeated experiments and sweeping of parameters without software intervention.

The readout time is determined by the user in Python and the readout is triggered by the tProcessor. The readout pulse offset and length is also configured by the user in Python.

The readout block can implement quantum feedback protocols. An extra AXI stream output is added, which is updated immediately after the average operation.  This output is connected to one of the input ports of the tProcessor.  The tProcessor can read IQ data from this external port, which can then be used in a conditional branch statement. 

\section{Performance characterization} \label{RFpnm}

\begin{figure*}[]
    \centering
    \includegraphics[width= 0.99\textwidth]{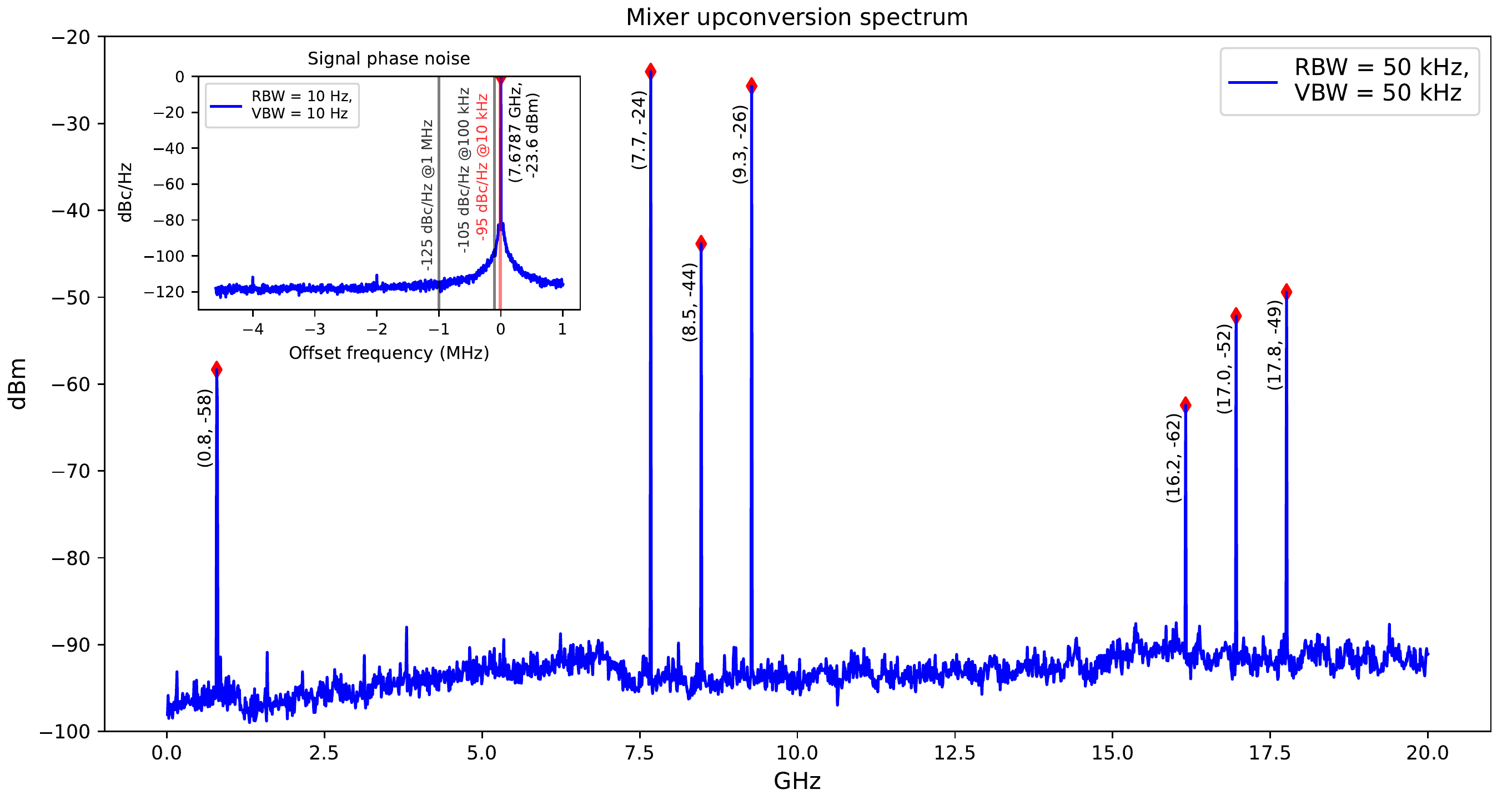} %
    \caption{QICK RF board performance: spectrum of mixer upconversion from IF to RF; inset is signal phase noise. In this measurement the RF board's additional $>$8 GHz low-pass filter was disabled so as to display the full upconversion spectrum. Here an 0.8 GHz IF has been upconverted with an on-board LO set to 8.478 GHz. The mixer spectrum is clean and shows the expected sideband spurs at IF, LO$\pm$IF, and 2*LO$\pm$IF. We measure $-95$ dBc/Hz of phase noise at 10 kHz offset from the carrier. Table \ref{tab:phasenoise} summarizes the phase noise measurements.}%
    \label{fig:rfperformance}%
\end{figure*}

Here we discuss the RF performance of the signal generator output. Fig.~\ref{fig:upconversion} shows the schematic of mixer upconversion from IF to RF using the QICK RF board. Since the digitally generated IF can cover 3 GHz of bandwidth in each of the first and second Nyquist zones, we recommend parking the LO at a frequency between 7.5 and 8.5 GHz. The RF board has a low-pass filter that suppresses frequencies over 8 GHz by more than 30 dB of attenuation. We use the lower sideband of the mixer for signal placement. With the LO at 7.5 to 8.5 GHz, the user can place pulses in a 3 GHz band without generating harmonics, and having enough LO suppression. Fig.~\ref{fig:rfperformance} shows a demonstration of the clean mixer upconversion spectrum (with the RF board's additional $>$8 GHz low-pass filter removed). Here an 0.8 GHz IF has been upconverted with an on-board LO set to 8.478 GHz. From 0.5 MHz to 8 GHz the mixer spurs are smaller than $-60$ dBc. Using the LO at lower frequencies is totally viable but the LO feedthrough will be seen at a power of $\sim-50$ dBm. An external low-pass filter or band-pass filter can be added to eliminate the LO feedthrough.

The QICK RF board has been designed to avoid the generation of undesired spurs in the working band of interest and to avoid lengthy calibrations that drift during the course of a single-qubit experiment. One of the main reasons why RF electronics require calibration is that analog IQ mixers have unequal complex gains. Analog IQ mixers are made of two mixers connected by a $90^\circ$ phase rotation. The amplitude and gain of the I and Q mixers over a large bandwidth (e.g. 4-8 GHz) typically differ by $\pm$ half a dB and few degrees. Calibrating the IQ mixer requires multiplying the signal by a frequency-dependent amplitude/phase 2$\times$2 matrix. The equations to calculate the matrix values are more involved for multiple tones. For instance, the mixer calibration process for MKID tones achieves sideband rejection of at best 30 dB and frequent recalibrations are required due to temperature drift. The QICK RF board avoids analog IQ mixers and their calibration requirements by instead using the high frequency DDS and digital IQ mixers in the FPGA, which are inherently balanced. There is no gain error other than a round-off bit that introduces less than $-100$ dBm of error. So, unlike most qubit controllers, the QICK does not require any mixer calibration.

The analog RF mixer on the QICK RF board is a non-IQ DSM (double sideband mixer). The DSM generates the two sidebands at the frequencies LO$\pm$IF. The mixer has some direct IF and LO leakage into the RF output, and the mixer nonlinearity generates spurs at the frequencies $n$LO $\pm$ $m$IF for integers $n$ and $m$. Fig.~\ref{fig:upconversion} shows how the QICK RF board filters all undesired outputs. As mentioned in Section \ref{signalgenerator} the fast DDS allows the user to place an IF pulse anywhere in a 3 GHz wide spectrum (the left rectangle in Fig.~\ref{fig:upconversion} before the mixer symbol). For instance, as in Fig.~\ref{fig:upconversion} the $\pm$3 GHz IF spectrum could be shifted up by an 8.5 GHz on-board LO and the two sidebands would be at 5-8 GHz and 9-12 GHz. The IF and LO feedthrough, and  $n$LO $\pm$ $m$IF products do not fall inside the 5-8 GHz band and are easily filtered. We have 500 MHz of room left between the LO and the LSB of interest, which allows for 35+ dB of filtering. Moving the LO in the 7.5 to 8.5 GHz band enables us to place our upconverted signal in the 4 to 8 GHz band. Fig.~\ref{fig:rfperformance} shows the unfiltered mixer output using a 8.478 GHz LO and 800 MHz IF. A typical 17 dB LO is attenuated by 66 dB (to $-48$ dBm). When the LO is placed at 8.478 GHz , the Mini-Circuits LFCW-6000+ adds another 45 dB of attenuation (to $-93$ dBm). The noise floor in the 4-8 GHz band is $-135$ dBc/Hz. The phase noise near the carrier is shown in the inset of  Fig.~\ref{fig:rfperformance}. Table \ref{tab:phasenoise} shows phase noise measurements from 100 Hz-10 MHz. 

The QICK's measured phase noise of $-95$ dBc/Hz at 7.6787 GHz carrier at 10 kHz offset is comparable to that of commercial qubit controllers which use direct digital synthesis (e.g. the Zurich Instruments SHFQA has a measured phase noise of $-96$ dBc/Hz at 8 GHz carrier at 10 kHz offset \cite{SHFQA}). For the highly dynamic pulse generation necessary for most qubit control applications, the QICK measured phase noise is sufficient for low gate error operations. For instance, it is below the value specified in \cite{VanDijk2020} for >99.9\% fidelity single-qubit control ($-116$ dBc/Hz at a 1 MHz oﬀset from the carrier). Also, one can compute its corresponding qubit dephasing spectral density using the method in \cite{Ball2016} and find it to generally be several orders of magnitude lower than that of the lab-grade local oscillator studied in that paper. For applications which require ultra-low phase noise, the user can substitute the QICK RF board with their choice of upconversion circuit whose dedicated local oscillator has lower phase noise. The QICK phase noise is limited by the RF mixer, since we measured the phase noise directly out of the RFSoC DAC to be lower by 15 to 20 dB. In future RFSoC generations, faster DACs which entirely eliminate the use of IQ mixers may help to reduce phase noise.

\begin{table}[h]
\small
\begin{center}
\begin{tabular}{|c | c|} 
 \hline
 Delta freq. (Hz) & Phase noise (dBc/Hz)\\
 \hline
 100 & -70 \\
 \hline
 1k & -80 \\
 \hline
 10k & -95 \\
 \hline
 100k & -105 \\
 \hline
 1M & -125 \\
 \hline
 10M & -125 \\
 \hline
\end{tabular}
\caption{\label{tab:phasenoise} QICK RF board phase noise measurements. In these measurements the carrier tone frequency was 7.6787 GHz.}
\end{center}
\end{table}

We also measured gain stability of the QICK versus time for a total of 16 hours. In this test, the controller played a single tone in loopback mode and the gain was measured in ADC digital units. No calibration was performed during this time besides the default background gain and temperature calibration that occurs in the RFSoC ADC \cite{PG269}. The board was sitting at room temperature with no additional temperature control. Over that time period the QICK measured gain had an RMS value of 0.07\%, which is sufficient for low-gate-error operations. For instance, this analog gain stability surpasses the value of 0.22\% specified in \cite{VanDijk2020} for >99.9\% fidelity single-qubit control.  

\subsection{Latency} \label{Latency}
\begin{figure}[h!]
\begin{center}
\includegraphics[width= 1\columnwidth]{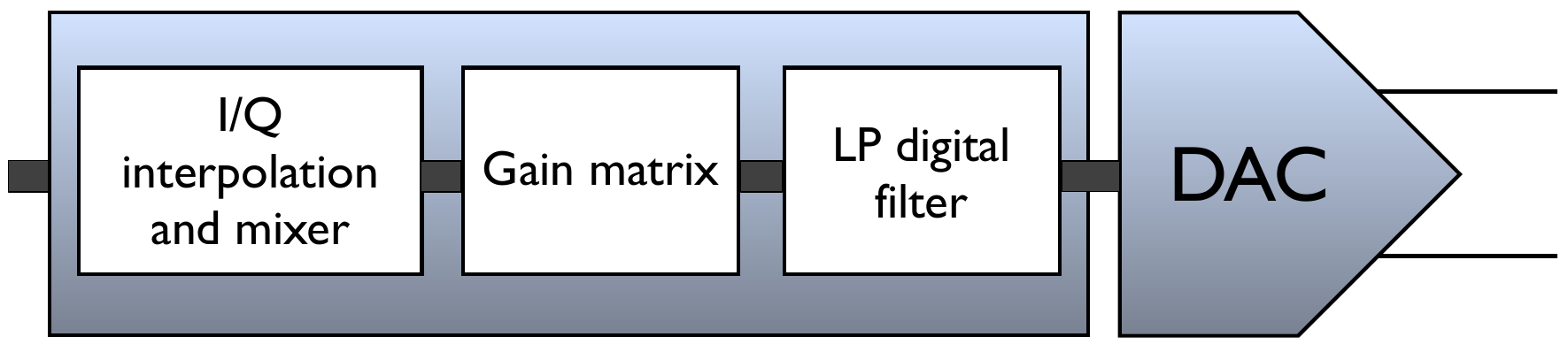}
        \caption{Simplified RFSoC DAC module block diagram. The ADC block diagram is similar but the interpolator block is replaced by a decimator block.}
        \label{fig:DAC_bd}
\end{center}
\end{figure}

Low latency is important for running deep circuits on noisy qubits that decohere quickly. If the qubit experiment is conditional upon the previous value read out from the system, the readout latency must also be included. Table \ref{tab:latencies} details measured latency numbers for the QICK. As mentioned in Section \ref{Intro}, the RFSoC DAC and ADC modules have a cascade of processing blocks that can be used or bypassed. The latency depends on the number of blocks being used. Fig.~\ref{fig:DAC_bd} shows a simplified block diagram of the DAC. The ADC is very similar, but the interpolation is replaced by a decimation. The round trip latency including the DAC, ADC, and the digital AXI interfaces associated to the DAC and ADC, was measured using an ILA (Internal Logic Analyzer) running at 512 MHz. When we bypass the interpolation/decimation and filters, the latency is 90 ns (46 ILA clocks). Half of that latency is due to the DAC and half to the ADC. When the upconversion and downconversion are enabled, the round trip latency increases to 113 ns (58 ILA clocks). When the x8 interpolation/decimation and filters are included, the latency goes up to 117 ns (60 ILA clocks) round trip. The loopback between DAC and ADC was made with a short coax cable that was a few inches long.

Another test was created to measure the latency of the logic in feedback readout mode. A readout pulse was output and fed-back into an ADC while a second DAC and a marker were used to measure times. The experiment shows that the conditional evaluation and address jump latency is 16 clock cycles. At the current firmware ADC sample rate, that totals 42 ns. If an address jump occurs the next pulse latency is 20 clock ticks, which equals 52 ns. Summing the measurements, the total latency of the QICK is 184 ns to 211 ns depending on the DAC and ADC configuration. A new version of the tProcessor is under development using pipelining to reduce the logic latency due to internal processing and instruction branching.

\begin{table}[h!]
\small
\begin{center}
\begin{tabular}{|m{1.7cm} | m{3.4cm} | c | c|} 
 \hline
 Sampling freq. (MHz) & Functions & Latency \\
 \hline
 ADC:4096
 DAC:6144 & ADC and DAC with all digital features bypassed & 90 ns \\
 \hline
 ADC:4096
 DAC:6144  & ADC and DAC with NCO enabled & 113 ns \\
 \hline
 ADC:4096
 DAC:6144  & ADC and DAC, NCO enabled, and interpolation/decimation enabled & 117 ns \\
 \hline
 ADC:3072
 DAC:6144   & Conditional evaluation and address jump & 42 ns \\
 \hline
 ADC:3072
 DAC:6144  & Next pulse latency if address jump=TRUE & 52 ns \\
 \hline
\end{tabular}

  \caption{\label{tab:latencies} Latency measurements taken to help develop the QICK firmware. Note that neither the NCO or the x8 interpolation/decimation and filters are enabled in the current version of the QICK firmware, although they may be used in future firmware versions.}
\end{center}
\end{table}

\subsection{Characterization of a transmon qubit}\label{sec:qubit}

\begin{figure}[h!]
    \centering
    \subfloat[]{{\includegraphics[width=1\columnwidth]{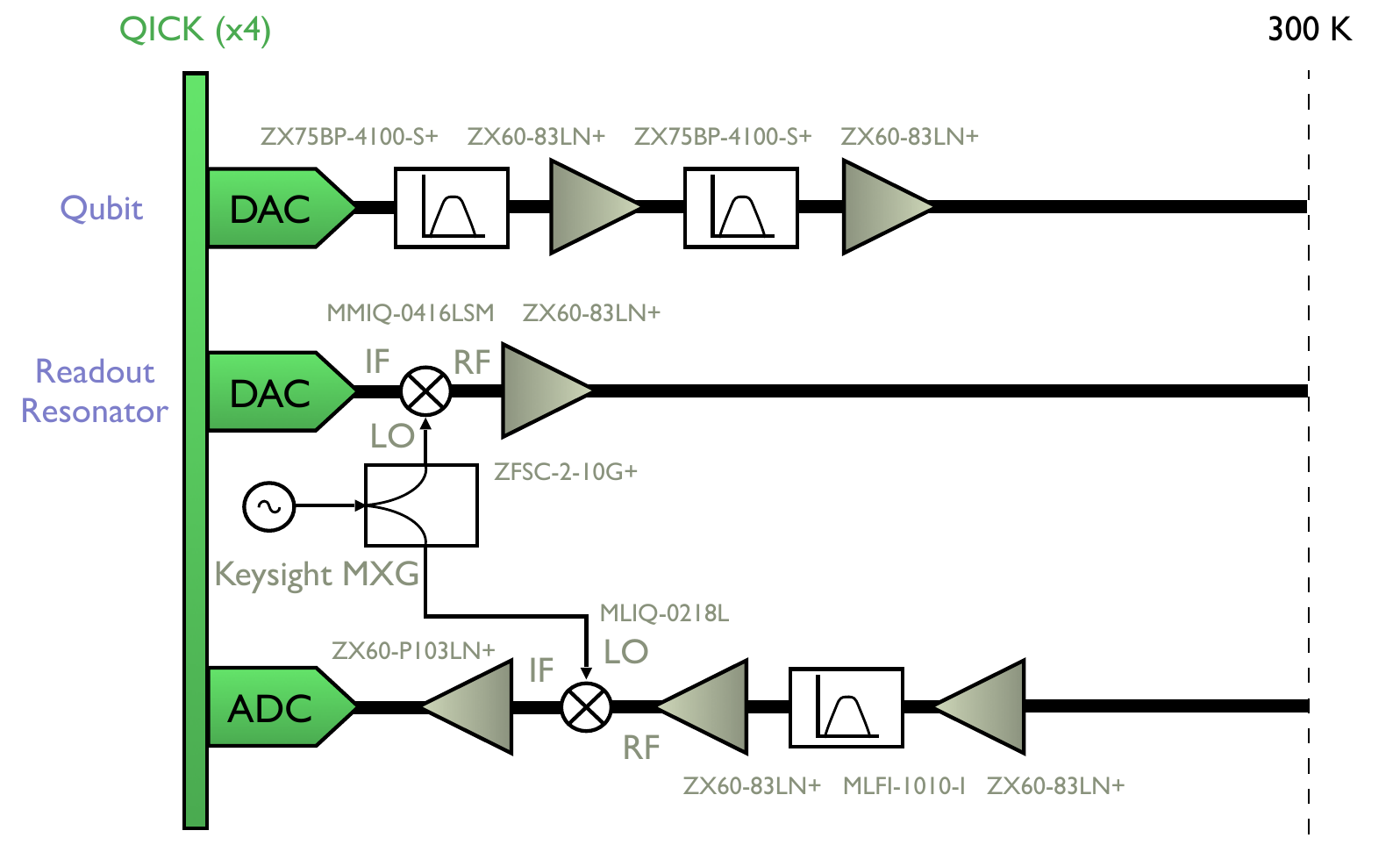} }}%
    \\
    \subfloat[]{{\includegraphics[width=1\columnwidth]{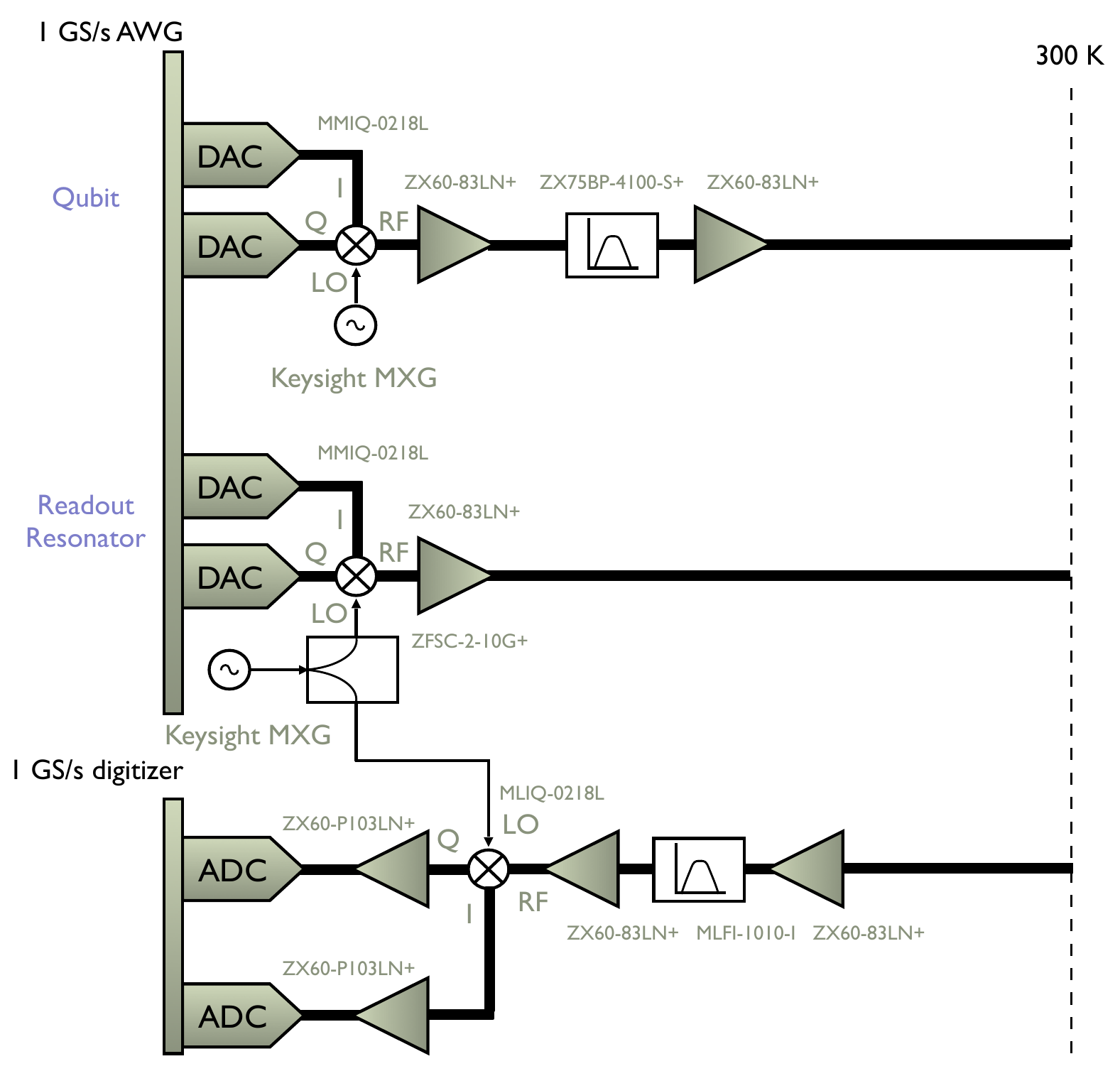} }}%
    \caption{Comparison of wiring diagrams with the QICK ZCU111 evaluation board (a) and with a conventional AWG setup (b). The QICK has eight RF DACs which can be configured in multiple ways. Using the QICK, two RF DACs are needed to control the qubit and its readout resonator, respectively. Also, the qubit control pulses (< 6 GHz) can be directly synthesized without the use of an analog mixer. This setup was used to gather the data shown in Figure \ref{fig:transmon_meas}. Note that the QICK RF board was not used because it was still under development. In a conventional AWG setup, four DACs would be needed to control a qubit and its readout resonator, and both the qubit and the readout resonator pulses would need to be upconverted with analog mixers.}
    \label{fig:wiring}
\end{figure}

\begin{figure*}[!ht]
    \centering
    \includegraphics[width= 0.99\textwidth]{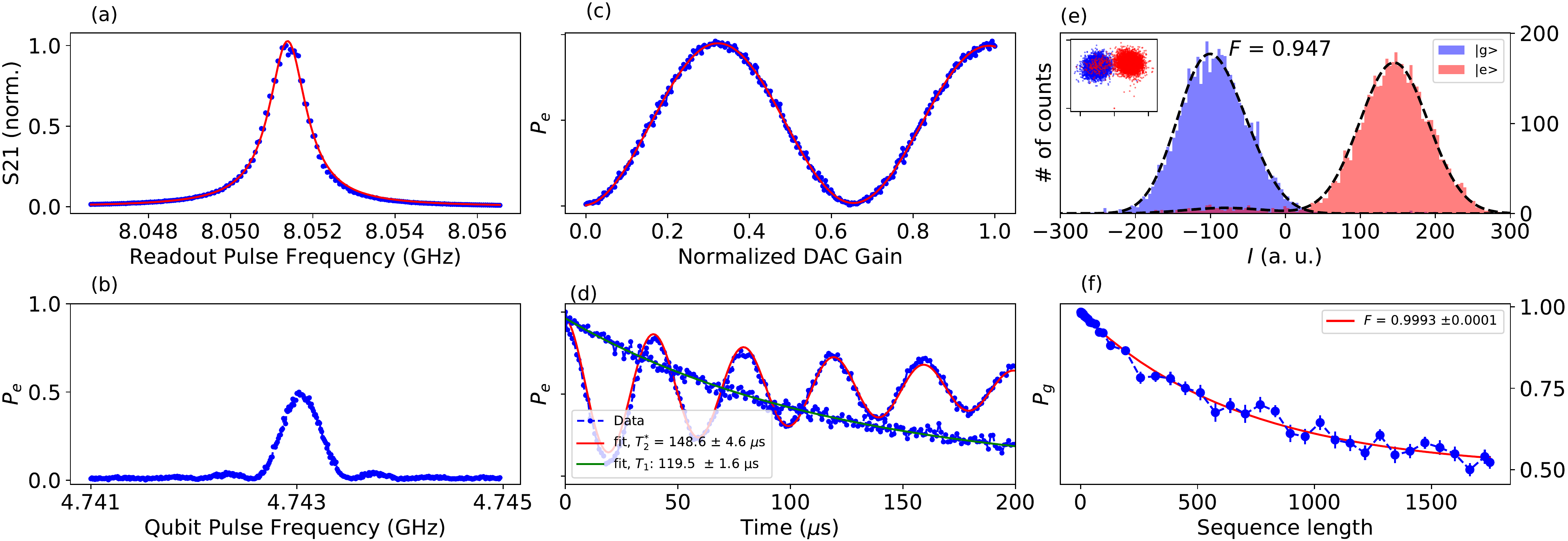}
    \caption{The QICK was used to characterize a transmon qubit dispersively coupled to a readout cavity. Figures (a) and (b) show readout cavity and qubit spectroscopy measurements, and figure (c) shows qubit Rabi oscillations. Figure (d) show qubit $T_1$ and $T_2$ measurements (of 119 $\mu$s and 148 $\mu$s, respectively). Figure (e) (inset) shows single shot digitized values when the qubit is prepared in ground and excited states. Histograms were fitted with a bi-modal Gaussian distribution function, resulting in 94.7\% fidelity with no parametric amplification. Figure (f) shows a randomized benchmarking protocol averaged over 30 unique random sequences with 3000 shots per data point. The average gate fidelity is $\mathcal{F}_{avg}=99.93\% \pm 0.01\%$ which approaches the estimated coherence-limited gate fidelity of $\mathcal{F}_{lim}=99.96\%$. All measurements were performed entirely on the QICK, by running a Jupyter notebook on-board which compiled experiments and acquired data.}%
    \label{fig:transmon_meas}
\end{figure*}

Describing superconducting qubit dynamics largely exceeds the scope of this paper. There are hundreds of excellent papers on the subject such as \cite{Koch2007, Gambetta2006, Nakamura1999, Fowler2012, Wallraff2005, Martinis2009, Preskill1997, DiVincenzo2000, Schoelkopf2008, Devoret2013}. In this section we will summarize only the information necessary to explain the functionality of the QICK. 

The classical control and readout of superconducting qubits allows for information to travel from the classical world to the quantum world and vice versa. High-quality classical control enables both a) the preparation of large coherent quantum states composed of entangled qubits and b) the manipulation and readout of these large quantum states. Such tools are integral to the execution of algorithms such as error correction and factoring, as well as to the preparation and measurement of states for quantum simulation.

Superconducting qubit systems are susceptible to noise-induced decoherence, and faulty qubit control (for instance, driving the qubit with spurious frequency components) can lead to driving unwanted transitions that cause the system to decohere. Noisy qubit readout is also a source of faulty control. Typically the readout noise is determined by the first amplifier in the readout chain (a cryogenic LNA such as a HEMT) and not by the warm electronics. Therefore, the QICK RF board was designed to ensure that the input noise of the warm electronics would be less than the cryogenic LNA output noise to avoid lowering the signal to noise ratio (SNR) of the readout signal. Also, the QICK RF board's efficient filtering of the warm readout electronics was designed to enhance the readout signal's SNR. 

Superconducting qubits are typically coupled to a resonant RF cavity. The qubit state is read out using a quantum non-demolition (QND) technique in which the state of the qubit is projected onto the RF cavity \cite{Schuster2007}. Repeated QND measurements of the qubit yield the same result. The readout electronics is broadband (i.e. several GHz), therefore several qubits, typically 100 MHz bandwidth, can be frequency multiplexed and read out simultaneously. A typical readout scheme measures transmitted power $S_{21}$ through the cavity that is coupled to the qubits. 

The high output bandwidth (4 GHz) of the QICK RF DACs and RF ADCs 1) improves the overall quality of qubit control by enabling the direct synthesis of spectrally pure qubit control pulses in the second Nyquist zone and 2) allows the controller to multiplex qubit control and readout. The multiplexing functionality is possible with the current QICK system but has not yet been implemented in an experiment. Our eventual aim is to multiplex up to 20 qubits (with 100 MHz bandwidth and 200 MHz separation) on the same control line, increasing the number of qubits to 100+ per board and several thousand in a modest system with few tens of boards.

The digital part of the QICK (loaded onto the ZCU111 evaluation board with the standard Xilinx XM500 RFMC balun card) was deployed and used to control qubits in the Schuster lab at the University of Chicago's James Franck Institute and Pritzker School of Molecular Engineering. The controller took high-quality single-qubit data even without the custom RF board, which was still being developed at Fermilab at the time the measurements were taken. The custom RF board was replaced by connectorized amplifiers, attenuators, mixers and filters which did the necessary amplification and up/downconversion. The overall controller performance was found to be on par with commercial qubit controllers that are ten times as expensive. The QICK controller was also straightforward for researchers to use, particularly due to its ability to directly synthesize carrier frequencies of up to 6 GHz. Compared to the conventional qubit controller used in the Schuster lab, the QICK required half the DAC channels and half the amount of analog upconversion (Figure \ref{fig:wiring}). The room-temperature and cryogenic setup originally used to measure this qubit before the QICK was introduced is shown in Figure S2 of \cite{Dixit2021}.

The QICK was used to control and read out a 3D transmon qubit that is being used as a dark matter detector \cite{Dixit2021}. The readout resonator pulses were generated and upconverted using a mixer with an IF frequency of 100 MHz. The IF frequency was swept to perform a resonator spectroscopy as shown in Fig.~\ref{fig:transmon_meas} (a). The qubit pulses  were directly synthesized operating one of the DAC channels in the second Nyquist zone. The qubit probe frequency was swept around 4.743 GHz followed by a readout tone to obtain Fig.~\ref{fig:transmon_meas} (b). The amplitude or power Rabi measurement was performed by varying the pulse amplitude of a 100 ns ($\sigma$ = 25 ns) long Gaussian pulse (Fig.~\ref{fig:transmon_meas} (c)). Ramsey interferometry was performed by preparing the qubit in a superposition state with a  $\frac{\pi}{2}$ pulse followed by a variable delay ($\tau$) before applying another $\frac{\pi}{2}$ pulse with a phase advanced by $\Delta \phi = \omega * \tau$, where $\omega$ is the Ramsey frequency (Fig.~\ref{fig:transmon_meas} (d)). A $T_1$ measurement was performed by preparing the qubit in its excited state with a $\pi$ pulse followed by a variable delay ($\tau$) before measuring the qubit's state (Fig.~\ref{fig:transmon_meas} (d)). The single shot IQ readout values were acquired for qubit prepared in ground (blue dots) and excited state (red dots) to estimate the readout fidelity. The resulting distribution was fitted with a bi-modal Gaussian function to extract the single shot fidelity, $F = 94.7\%$ without any parametric amplification as shown in Fig. \ref{fig:transmon_meas} (e). Lastly, we characterize the fidelity of our single-qubit gates through a randomized benchmarking protocol. The gates were randomly chosen from this gate set \{I, X, Y, Z, X/2, -X/2, Y/2, -Y/2, Z/2, -Z/2\} where $Z$ and $\pm Z/2$ gates were implemented virtually by advancing the phase of subsequent gates. For each sequence length, we use 30 randomized sequences, each containing a recovery gate to bring the qubit back to its ground state before performing a readout. Each individual data point is averaged over 3000 shots. The average gate fidelity was found to be $\mathcal{F}_{avg}=99.93\% \pm 0.01\%$ as shown in Fig. \ref{fig:transmon_meas} (f). The coherence-limited gate fidelity is estimated to be $\mathcal{F}_{lim}=99.96\%$. The duration and integration time of the readout pulse for these measurements was 3 $\mu$s. This is related to the qubit-readout dispersive shift ($\frac{\chi}{2\pi} \sim 350$ kHz) which is limited by the device design and not by the control hardware. Interested readers may follow excellent review articles describing standard qubit characterization experiments \cite{Oliver, Mahdi}.  

Controlling this transmon with the QICK was not found to degrade its coherence relative to the coherences measured with two different commercial systems: a conventional Keysight AWG-based system and the Quantum Machines OPX system. The $T_1$ decay time measured with the QICK was 119.5 $\pm$ 1.6 us (versus 108 $\pm$ 18 us measured with the AWG-based system \cite{Dixit2021} and 122.6 $\pm$ 2.5 $\mu$s with the OPX). The $T_2$ decay time measured with the QICK was 148.6 $\pm$ 4.6 $\mu$s (versus 61 $\pm$ 4 $\mu$s measured with the AWG-based system and 155.9 $\pm$ 3.3 $\mu$s with the OPX). The AWG-based system measured a significantly lower $T_2$ because there was increased readout population for that cooldown, which occurred prior to the cooldown where the QICK and the OPX were each used to measure the device. The remaining discrepancies between these values can be attributed to the well-documented phenomenon of superconducting qubit coherence fluctuating over long time scales \cite{Place2021}\cite{carroll2021dynamics}. Additionally, the readout fidelity measured with the QICK was 94.7\%, which is comparable to the two other measurements of readout fidelity (95.8 $\pm$ 0.4\% measured with the AWG-based system and 95\% measured with the OPX). The remaining discrepancies between these values can be attributed to differences in the readout pulse length and envelope shape as well as differences in digitization methods used for the respective experiments conducted.

The QICK has also been deployed in the Schuster lab to benchmark heavy fluxonium qubits with fast flux pulses instead of standard RF microwave pulses. The lab recently demonstrated this technique in \cite{Zhang2021} with a conventional control setup and was afterwards able to reproduce their results with the QICK. Such experiments require the QICK DACs to be in DC coupled mode so that they can generate fast unmodulated pulses.

\section{Summary and future work}

We introduced an RFSoC-based qubit controller called the Quantum Instrumentation Control Kit (QICK) that is capable of directly synthesizing qubit control pulses with IF frequencies up to 3 GHz (6 GHz) in the first (second) Nyquist zone. The QICK can be used to control multi-qubit systems. The low cost of the controller, roughly \$15,000 USD for eight RF DAC channels and eight RF ADC channels, makes it useful for scaling up qubit experiments in academic laboratories.

Despite the short development time (1.5 years), our team has added low and high-level functionalities to operate complex qubit systems and experiments. The QICK provides an integrated solution for qubit biasing, control and readout. All RF and synchronization devices are included on-board. Users can freely access the QICK Github repository with controller firmware, software, and documentation \cite{QICKrepo}. 

Our future work focuses on hardware, firmware and software improvement and improved functionality. On the hardware side, we are exploring the use of the Xilinx ZCU216 board hosting a ZU49DR RFSoC Gen3 FPGA \cite{zcu216}. The ZU49DR DACs run at 10 GS/s, with increased analog bandwidth, > 6 GHz. Many more typical qubit experiments could now be run without the need of an external analog mixer, further simplifying the setup. The ZU49DR has 16 DACs and 16 ADCs, which implies scaled-up firmware functionality and throughput. To allow for faster qubit experiments with more channels we will increase the number of tProcessor microsequencers in the FPGA. We will allocate one microsequencer per channel or per every few channels without compromising FPGA resources. To improve the flexibility of the signal generator we will 1) integrate the variable length and interpolation envelope engines, which are currently in a beta phase and 2) develop drivers for very long AWG pulses requiring DDR4 and interpolation. We will also develop and implement optimal filters for readout.

Much of our future work focuses on the software and software-firmware interface. We will integrate the QICK with a simulator and debugger to cut the qubit experiment program development time. We will also integrate our current software with high-level software packages such as Qiskit and OpenQASM \cite{qiskit, openqasm}, which will help developers.

\section{Acknowledgements}
This manuscript has been authored by Fermi Research Alliance, LLC under Contract No. DE-AC02- 07CH11359 with the U.S. Department of Energy, Office of Science, Office of High Energy Physics, with support from its QuantISED program and from National Quantum Information Science Research Centers, Quantum Science Center. This work was funded in part by EPiQC, an NSF Expedition in Computing, under grant CCF-1730449. This work was supported by the Army Research
Office under Grant No. W911NF1910016. S Sussman is supported by the Department of Defense (DoD) through the National Defense Science \& Engineering Graduate Fellowship (NDSEG) Program. A Agrawal is supported by the Heising-Simons Foundation. Salvatore Montella is supported by DOE SQMS and from the Quantum Science Center and Superconducting Quantum Materials and Systems Center. The Fermilab team thanks Gaston Gutierrez (Fermilab) for his help bridging the gap between engineering and quantum mechanics. The authors thank National Quantum Information Science Research Centers Q-NEXT, SQMS under contract number DE-AC02-07CH11359, and C2QA members who participated in discussions.

\section{Availability of data}
The data and artwork that support the findings of this study are available from the corresponding author upon reasonable request.

\begin{appendices}

\section{Two-board synchronization}
\label{Two board synchronization}
\begin{figure}[h!]
\begin{center}
\includegraphics[width=1\columnwidth]{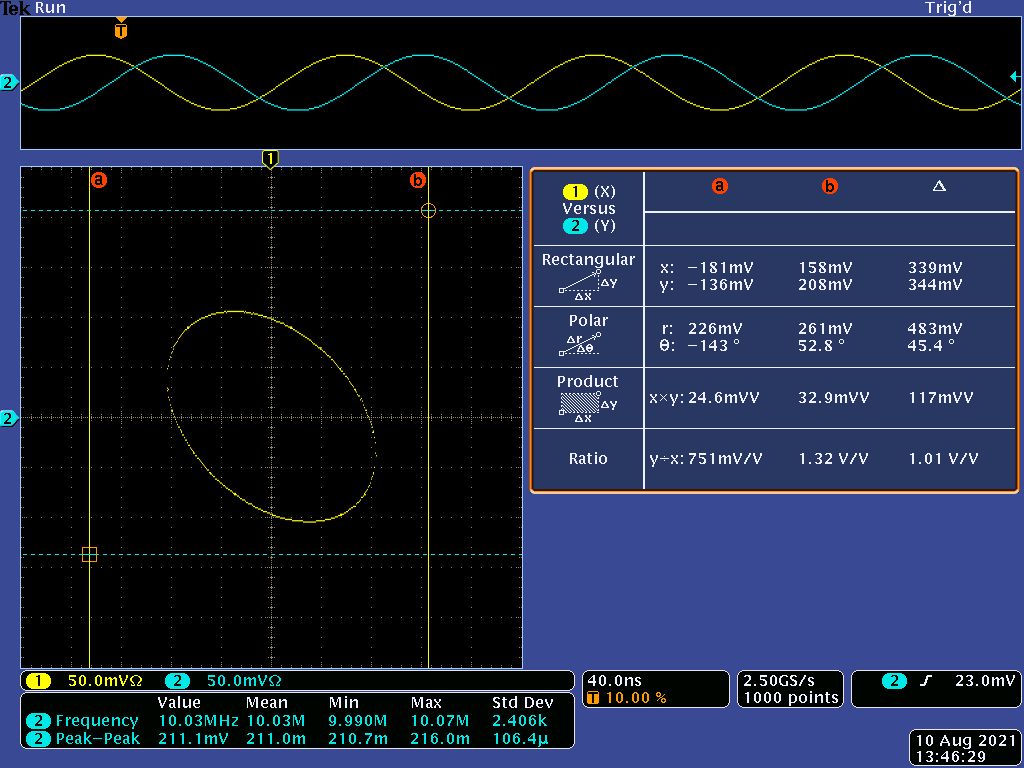}
        \caption{Lissajous figure from two-board synchronization. The output of a DAC from one ZCU111 was used to drive the horizontal input of a Tektronix oscilloscope and a DAC output from the other ZCU111 was used to drive the vertical scope input.}
        \label{fig:two_board_sync}
\end{center}
\end{figure}

Two ZCU111 boards were synchronized to an external rubidium stable source. On each board a DAC output was used to generate a continuous single 10 MHz tone using a fast DDS and a constant envelope. The output of a DAC from one ZCU111 was used to drive the horizontal input of a Tektronix oscilloscope and a DAC output from the other ZCU111 was used to drive the vertical scope input. Figure \ref{fig:two_board_sync} shows the typical Lissajous figure. The plot remaines stable independently of time (e.g. hours). Turning the synchronization between the two boards off and on again only modifies the initial phase of the Lissajous figure, which remains stable.

\newpage

\section{Bandwidth table}
\label{bandwidths} 

\begin{table}[h!]
\small
\begin{center}
\begin{tabular}{|m{2.2cm} | m{2.2cm} | m{3cm} | c|} 
 \hline
 Device & Frequency (GHz) & Comments \\
 \hline
  Digital DDS & 0 to 3,  3 to 6 & $1^{st}, 2^{nd}$ Nyquist zones\\
 \hline
 DAC NCO & 0 to $F_s$ & Not used in current version of firmware. \\
 \hline
 DAC analog BW & 6 & Measured using broadband white Gaussian noise \\
 \hline
 Analog mixer LO frequency & 2-15 & Tunable range \\
 \hline
 RF low-pass filter  & 1.8 & LFCN-1800+ fixed but could be replaced \\
\hline
 RF high-pass filter  & 6.3 &  LFCW-6300+ fixed but could be replaced \\
 \hline
\end{tabular}

  \caption{Bandwidths of different aspects of the QICK system. NOTE: the first three lines in the table reflect the allowable digital and analog BW of the RFSoC board. the last three lines reflect the analog BW on the QICK RF custom board. The filters on the RF board have been chosen to minimize RF noise of current experiments. Those filters can be replaced to increase or reduce the analog BW as needed.}
\end{center}
\end{table}

\section{The QICK Architecture Timelines}
\label{Timelines}
There are two different timelines in the QICK architecture: the tProcessor timeline of decision-making about the placement of events according to the master clock, and the master clock timeline itself. These timelines are decoupled although not totally independent since the tProcessor has to be ahead of the master clock timeline to avoid the master clock timeline being under-run and waiting for instructions that arrive too late. The tProcessor timeline is managed as follows:

\begin{enumerate}[i]
  \item Compiled quantum algorithm instructions are written into the tProcessor memory.
  \item The tProcessor decodes instructions. If the instruction can be executed immediately (e.g. queue a pulse generator request, configure hardware or firmware parameters) it does so.
  \item The tProcessor waits for specific events such as pulse generator queues being full, channel readout triggers, feedback from a readout, or a forced wait.
\end{enumerate}

The master clock timeline is managed as follows:

\begin{enumerate}[i]
  \item All signal generator pulses are played at the exact time as specified by the time in the queue.
  \item Every signal generator has its own queue and they are all synchronized to the same master clock. More than one signal generator may output a pulse at the same time or overlapping time with other channels.
  \item Readout channels, upon a tProcessor trigger, are synchronized to the master clock.
\end{enumerate}

\end{appendices}

\bibliography{bibliography1.bib}

\providecommand{\noopsort}[1]{}\providecommand{\singleletter}[1]{#1}%
\begin{thebibliography}{59}%
\makeatletter
\providecommand \@ifxundefined [1]{%
 \@ifx{#1\undefined}
}%
\providecommand \@ifnum [1]{%
 \ifnum #1\expandafter \@firstoftwo
 \else \expandafter \@secondoftwo
 \fi
}%
\providecommand \@ifx [1]{%
 \ifx #1\expandafter \@firstoftwo
 \else \expandafter \@secondoftwo
 \fi
}%
\providecommand \natexlab [1]{#1}%
\providecommand \enquote  [1]{``#1''}%
\providecommand \bibnamefont  [1]{#1}%
\providecommand \bibfnamefont [1]{#1}%
\providecommand \citenamefont [1]{#1}%
\providecommand \href@noop [0]{\@secondoftwo}%
\providecommand \href [0]{\begingroup \@sanitize@url \@href}%
\providecommand \@href[1]{\@@startlink{#1}\@@href}%
\providecommand \@@href[1]{\endgroup#1\@@endlink}%
\providecommand \@sanitize@url [0]{\catcode `\\12\catcode `\$12\catcode
  `\&12\catcode `\#12\catcode `\^12\catcode `\_12\catcode `\%12\relax}%
\providecommand \@@startlink[1]{}%
\providecommand \@@endlink[0]{}%
\providecommand \url  [0]{\begingroup\@sanitize@url \@url }%
\providecommand \@url [1]{\endgroup\@href {#1}{\urlprefix }}%
\providecommand \urlprefix  [0]{URL }%
\providecommand \Eprint [0]{\href }%
\providecommand \doibase [0]{http://dx.doi.org/}%
\providecommand \selectlanguage [0]{\@gobble}%
\providecommand \bibinfo  [0]{\@secondoftwo}%
\providecommand \bibfield  [0]{\@secondoftwo}%
\providecommand \translation [1]{[#1]}%
\providecommand \BibitemOpen [0]{}%
\providecommand \bibitemStop [0]{}%
\providecommand \bibitemNoStop [0]{.\EOS\space}%
\providecommand \EOS [0]{\spacefactor3000\relax}%
\providecommand \BibitemShut  [1]{\csname bibitem#1\endcsname}%
\let\auto@bib@innerbib\@empty
\bibitem [{\citenamefont {Shor}(1997)}]{Shor_1997}%
  \BibitemOpen
  \bibfield  {author} {\bibinfo {author} {\bibfnamefont {P.~W.}\ \bibnamefont
  {Shor}},\ }\href {\doibase 10.1137/s0097539795293172} {\bibfield  {journal}
  {\bibinfo  {journal} {SIAM Journal on Computing}\ }\textbf {\bibinfo {volume}
  {26}},\ \bibinfo {pages} {1484–1509} (\bibinfo {year} {1997})}\BibitemShut
  {NoStop}%
\bibitem [{\citenamefont {Gidney}\ and\ \citenamefont
  {Eker{\aa}}(2021)}]{Gidney2021}%
  \BibitemOpen
  \bibfield  {author} {\bibinfo {author} {\bibfnamefont {C.}~\bibnamefont
  {Gidney}}\ and\ \bibinfo {author} {\bibfnamefont {M.}~\bibnamefont
  {Eker{\aa}}},\ }\href {\doibase 10.22331/Q-2021-04-15-433} {\bibfield
  {journal} {\bibinfo  {journal} {Quantum}\ }\textbf {\bibinfo {volume} {5}},\
  \bibinfo {pages} {1} (\bibinfo {year} {2021})},\ \Eprint
  {http://arxiv.org/abs/1905.09749} {arXiv:1905.09749} \BibitemShut {NoStop}%
\bibitem [{\citenamefont {Bennett}\ and\ \citenamefont
  {Brassard}(2014)}]{Bennett_2014}%
  \BibitemOpen
  \bibfield  {author} {\bibinfo {author} {\bibfnamefont {C.~H.}\ \bibnamefont
  {Bennett}}\ and\ \bibinfo {author} {\bibfnamefont {G.}~\bibnamefont
  {Brassard}},\ }\href {\doibase 10.1016/j.tcs.2014.05.025} {\bibfield
  {journal} {\bibinfo  {journal} {Theoretical Computer Science}\ }\textbf
  {\bibinfo {volume} {560}},\ \bibinfo {pages} {7–11} (\bibinfo {year}
  {2014})}\BibitemShut {NoStop}%
\bibitem [{\citenamefont {Yin}\ \emph {et~al.}(2020)\citenamefont {Yin},
  \citenamefont {Li}, \citenamefont {Liao}, \citenamefont {Yang}, \citenamefont
  {Cao}, \citenamefont {Zhang}, \citenamefont {Ren}, \citenamefont {Cai},
  \citenamefont {Liu}, \citenamefont {Li}, \citenamefont {Shu}, \citenamefont
  {Huang}, \citenamefont {Deng}, \citenamefont {Li}, \citenamefont {Zhang},
  \citenamefont {Liu}, \citenamefont {Chen}, \citenamefont {Lu}, \citenamefont
  {Wang}, \citenamefont {Xu}, \citenamefont {Wang}, \citenamefont {Peng},
  \citenamefont {Ekert},\ and\ \citenamefont {Pan}}]{Yin2020}%
  \BibitemOpen
  \bibfield  {author} {\bibinfo {author} {\bibfnamefont {J.}~\bibnamefont
  {Yin}}, \bibinfo {author} {\bibfnamefont {Y.~H.}\ \bibnamefont {Li}},
  \bibinfo {author} {\bibfnamefont {S.~K.}\ \bibnamefont {Liao}}, \bibinfo
  {author} {\bibfnamefont {M.}~\bibnamefont {Yang}}, \bibinfo {author}
  {\bibfnamefont {Y.}~\bibnamefont {Cao}}, \bibinfo {author} {\bibfnamefont
  {L.}~\bibnamefont {Zhang}}, \bibinfo {author} {\bibfnamefont {J.~G.}\
  \bibnamefont {Ren}}, \bibinfo {author} {\bibfnamefont {W.~Q.}\ \bibnamefont
  {Cai}}, \bibinfo {author} {\bibfnamefont {W.~Y.}\ \bibnamefont {Liu}},
  \bibinfo {author} {\bibfnamefont {S.~L.}\ \bibnamefont {Li}}, \bibinfo
  {author} {\bibfnamefont {R.}~\bibnamefont {Shu}}, \bibinfo {author}
  {\bibfnamefont {Y.~M.}\ \bibnamefont {Huang}}, \bibinfo {author}
  {\bibfnamefont {L.}~\bibnamefont {Deng}}, \bibinfo {author} {\bibfnamefont
  {L.}~\bibnamefont {Li}}, \bibinfo {author} {\bibfnamefont {Q.}~\bibnamefont
  {Zhang}}, \bibinfo {author} {\bibfnamefont {N.~L.}\ \bibnamefont {Liu}},
  \bibinfo {author} {\bibfnamefont {Y.~A.}\ \bibnamefont {Chen}}, \bibinfo
  {author} {\bibfnamefont {C.~Y.}\ \bibnamefont {Lu}}, \bibinfo {author}
  {\bibfnamefont {X.~B.}\ \bibnamefont {Wang}}, \bibinfo {author}
  {\bibfnamefont {F.}~\bibnamefont {Xu}}, \bibinfo {author} {\bibfnamefont
  {J.~Y.}\ \bibnamefont {Wang}}, \bibinfo {author} {\bibfnamefont {C.~Z.}\
  \bibnamefont {Peng}}, \bibinfo {author} {\bibfnamefont {A.~K.}\ \bibnamefont
  {Ekert}}, \ and\ \bibinfo {author} {\bibfnamefont {J.~W.}\ \bibnamefont
  {Pan}},\ }\href {\doibase 10.1038/s41586-020-2401-y} {\bibfield  {journal}
  {\bibinfo  {journal} {Nature}\ }\textbf {\bibinfo {volume} {582}} (\bibinfo
  {year} {2020}),\ 10.1038/s41586-020-2401-y}\BibitemShut {NoStop}%
\bibitem [{\citenamefont {Wang}\ \emph {et~al.}(2020)\citenamefont {Wang},
  \citenamefont {Curtis}, \citenamefont {Lester}, \citenamefont {Zhang},
  \citenamefont {Gao}, \citenamefont {Freeze}, \citenamefont {Batista},
  \citenamefont {Vaccaro}, \citenamefont {Chuang}, \citenamefont {Frunzio},
  \citenamefont {Jiang}, \citenamefont {Girvin},\ and\ \citenamefont
  {Schoelkopf}}]{Wang_2020}%
  \BibitemOpen
  \bibfield  {author} {\bibinfo {author} {\bibfnamefont {C.~S.}\ \bibnamefont
  {Wang}}, \bibinfo {author} {\bibfnamefont {J.~C.}\ \bibnamefont {Curtis}},
  \bibinfo {author} {\bibfnamefont {B.~J.}\ \bibnamefont {Lester}}, \bibinfo
  {author} {\bibfnamefont {Y.}~\bibnamefont {Zhang}}, \bibinfo {author}
  {\bibfnamefont {Y.~Y.}\ \bibnamefont {Gao}}, \bibinfo {author} {\bibfnamefont
  {J.}~\bibnamefont {Freeze}}, \bibinfo {author} {\bibfnamefont {V.~S.}\
  \bibnamefont {Batista}}, \bibinfo {author} {\bibfnamefont {P.~H.}\
  \bibnamefont {Vaccaro}}, \bibinfo {author} {\bibfnamefont {I.~L.}\
  \bibnamefont {Chuang}}, \bibinfo {author} {\bibfnamefont {L.}~\bibnamefont
  {Frunzio}}, \bibinfo {author} {\bibfnamefont {L.}~\bibnamefont {Jiang}},
  \bibinfo {author} {\bibfnamefont {S.~M.}\ \bibnamefont {Girvin}}, \ and\
  \bibinfo {author} {\bibfnamefont {R.~J.}\ \bibnamefont {Schoelkopf}},\ }\href
  {\doibase 10.1103/PhysRevX.10.021060} {\bibfield  {journal} {\bibinfo
  {journal} {Phys. Rev. X}\ }\textbf {\bibinfo {volume} {10}},\ \bibinfo
  {pages} {021060} (\bibinfo {year} {2020})}\BibitemShut {NoStop}%
\bibitem [{\citenamefont {Biamonte}\ \emph {et~al.}(2017)\citenamefont
  {Biamonte}, \citenamefont {Wittek}, \citenamefont {Pancotti}, \citenamefont
  {Rebentrost}, \citenamefont {Wiebe},\ and\ \citenamefont
  {Lloyd}}]{Biamonte_2017}%
  \BibitemOpen
  \bibfield  {author} {\bibinfo {author} {\bibfnamefont {J.}~\bibnamefont
  {Biamonte}}, \bibinfo {author} {\bibfnamefont {P.}~\bibnamefont {Wittek}},
  \bibinfo {author} {\bibfnamefont {N.}~\bibnamefont {Pancotti}}, \bibinfo
  {author} {\bibfnamefont {P.}~\bibnamefont {Rebentrost}}, \bibinfo {author}
  {\bibfnamefont {N.}~\bibnamefont {Wiebe}}, \ and\ \bibinfo {author}
  {\bibfnamefont {S.}~\bibnamefont {Lloyd}},\ }\href {\doibase
  10.1038/nature23474} {\bibfield  {journal} {\bibinfo  {journal} {Nature}\
  }\textbf {\bibinfo {volume} {549}},\ \bibinfo {pages} {195–202} (\bibinfo
  {year} {2017})}\BibitemShut {NoStop}%
\bibitem [{\citenamefont {Egan}\ \emph {et~al.}(2021)\citenamefont {Egan},
  \citenamefont {Debroy}, \citenamefont {Noel}, \citenamefont {Risinger},
  \citenamefont {Zhu}, \citenamefont {Biswas}, \citenamefont {Newman},
  \citenamefont {Li}, \citenamefont {Brown}, \citenamefont {Cetina},\ and\
  \citenamefont {Monroe}}]{Egan_2021}%
  \BibitemOpen
  \bibfield  {author} {\bibinfo {author} {\bibfnamefont {L.}~\bibnamefont
  {Egan}}, \bibinfo {author} {\bibfnamefont {D.~M.}\ \bibnamefont {Debroy}},
  \bibinfo {author} {\bibfnamefont {C.}~\bibnamefont {Noel}}, \bibinfo {author}
  {\bibfnamefont {A.}~\bibnamefont {Risinger}}, \bibinfo {author}
  {\bibfnamefont {D.}~\bibnamefont {Zhu}}, \bibinfo {author} {\bibfnamefont
  {D.}~\bibnamefont {Biswas}}, \bibinfo {author} {\bibfnamefont
  {M.}~\bibnamefont {Newman}}, \bibinfo {author} {\bibfnamefont
  {M.}~\bibnamefont {Li}}, \bibinfo {author} {\bibfnamefont {K.~R.}\
  \bibnamefont {Brown}}, \bibinfo {author} {\bibfnamefont {M.}~\bibnamefont
  {Cetina}}, \ and\ \bibinfo {author} {\bibfnamefont {C.}~\bibnamefont
  {Monroe}},\ }\href@noop {} {\enquote {\bibinfo {title} {Fault-tolerant
  operation of a quantum error-correction code},}\ } (\bibinfo {year} {2021}),\
  \Eprint {http://arxiv.org/abs/2009.11482} {arXiv:2009.11482 [quant-ph]}
  \BibitemShut {NoStop}%
\bibitem [{\citenamefont {Kjaergaard}\ \emph {et~al.}(2020)\citenamefont
  {Kjaergaard}, \citenamefont {Schwartz}, \citenamefont {Braumüller},
  \citenamefont {Krantz}, \citenamefont {Wang}, \citenamefont {Gustavsson},\
  and\ \citenamefont {Oliver}}]{Kjaergaard_2020}%
  \BibitemOpen
  \bibfield  {author} {\bibinfo {author} {\bibfnamefont {M.}~\bibnamefont
  {Kjaergaard}}, \bibinfo {author} {\bibfnamefont {M.~E.}\ \bibnamefont
  {Schwartz}}, \bibinfo {author} {\bibfnamefont {J.}~\bibnamefont
  {Braumüller}}, \bibinfo {author} {\bibfnamefont {P.}~\bibnamefont {Krantz}},
  \bibinfo {author} {\bibfnamefont {J.~I.-J.}\ \bibnamefont {Wang}}, \bibinfo
  {author} {\bibfnamefont {S.}~\bibnamefont {Gustavsson}}, \ and\ \bibinfo
  {author} {\bibfnamefont {W.~D.}\ \bibnamefont {Oliver}},\ }\href {\doibase
  10.1146/annurev-conmatphys-031119-050605} {\bibfield  {journal} {\bibinfo
  {journal} {Annual Review of Condensed Matter Physics}\ }\textbf {\bibinfo
  {volume} {11}},\ \bibinfo {pages} {369–395} (\bibinfo {year}
  {2020})}\BibitemShut {NoStop}%
\bibitem [{\citenamefont {Yoneda}\ \emph {et~al.}(2017)\citenamefont {Yoneda},
  \citenamefont {Takeda}, \citenamefont {Otsuka}, \citenamefont {Nakajima},
  \citenamefont {Delbecq}, \citenamefont {Allison}, \citenamefont {Honda},
  \citenamefont {Kodera}, \citenamefont {Oda}, \citenamefont {Hoshi},\ and\
  \citenamefont {et~al.}}]{Yoneda_2017}%
  \BibitemOpen
  \bibfield  {author} {\bibinfo {author} {\bibfnamefont {J.}~\bibnamefont
  {Yoneda}}, \bibinfo {author} {\bibfnamefont {K.}~\bibnamefont {Takeda}},
  \bibinfo {author} {\bibfnamefont {T.}~\bibnamefont {Otsuka}}, \bibinfo
  {author} {\bibfnamefont {T.}~\bibnamefont {Nakajima}}, \bibinfo {author}
  {\bibfnamefont {M.~R.}\ \bibnamefont {Delbecq}}, \bibinfo {author}
  {\bibfnamefont {G.}~\bibnamefont {Allison}}, \bibinfo {author} {\bibfnamefont
  {T.}~\bibnamefont {Honda}}, \bibinfo {author} {\bibfnamefont
  {T.}~\bibnamefont {Kodera}}, \bibinfo {author} {\bibfnamefont
  {S.}~\bibnamefont {Oda}}, \bibinfo {author} {\bibfnamefont {Y.}~\bibnamefont
  {Hoshi}}, \ and\ \bibinfo {author} {\bibnamefont {et~al.}},\ }\href {\doibase
  10.1038/s41565-017-0014-x} {\bibfield  {journal} {\bibinfo  {journal} {Nature
  Nanotechnology}\ }\textbf {\bibinfo {volume} {13}},\ \bibinfo {pages}
  {102–106} (\bibinfo {year} {2017})}\BibitemShut {NoStop}%
\bibitem [{\citenamefont {Schirhagl}\ \emph {et~al.}(2014)\citenamefont
  {Schirhagl}, \citenamefont {Chang}, \citenamefont {Loretz},\ and\
  \citenamefont {Degen}}]{Schirhagl_2014}%
  \BibitemOpen
  \bibfield  {author} {\bibinfo {author} {\bibfnamefont {R.}~\bibnamefont
  {Schirhagl}}, \bibinfo {author} {\bibfnamefont {K.}~\bibnamefont {Chang}},
  \bibinfo {author} {\bibfnamefont {M.}~\bibnamefont {Loretz}}, \ and\ \bibinfo
  {author} {\bibfnamefont {C.~L.}\ \bibnamefont {Degen}},\ }\href {\doibase
  10.1146/annurev-physchem-040513-103659} {\bibfield  {journal} {\bibinfo
  {journal} {Annual Review of Physical Chemistry}\ }\textbf {\bibinfo {volume}
  {65}},\ \bibinfo {pages} {83} (\bibinfo {year} {2014})},\ \bibinfo {note}
  {pMID: 24274702},\ \Eprint
  {http://arxiv.org/abs/https://doi.org/10.1146/annurev-physchem-040513-103659}
  {https://doi.org/10.1146/annurev-physchem-040513-103659} \BibitemShut
  {NoStop}%
\bibitem [{\citenamefont {Bloch}(2005)}]{Bloch_2005}%
  \BibitemOpen
  \bibfield  {author} {\bibinfo {author} {\bibfnamefont {I.}~\bibnamefont
  {Bloch}},\ }\href {\doibase 10.1038/nphys138} {\bibfield  {journal} {\bibinfo
   {journal} {Nat Phys}\ }\textbf {\bibinfo {volume} {1}},\ \bibinfo {pages}
  {23} (\bibinfo {year} {2005})}\BibitemShut {NoStop}%
\bibitem [{\citenamefont {Sager}\ \emph {et~al.}(2020)\citenamefont {Sager},
  \citenamefont {Smart},\ and\ \citenamefont {Mazziotti}}]{Sager_2020}%
  \BibitemOpen
  \bibfield  {author} {\bibinfo {author} {\bibfnamefont {L.~M.}\ \bibnamefont
  {Sager}}, \bibinfo {author} {\bibfnamefont {S.~E.}\ \bibnamefont {Smart}}, \
  and\ \bibinfo {author} {\bibfnamefont {D.~A.}\ \bibnamefont {Mazziotti}},\
  }\href {\doibase 10.1103/physrevresearch.2.043205} {\bibfield  {journal}
  {\bibinfo  {journal} {Physical Review Research}\ }\textbf {\bibinfo {volume}
  {2}} (\bibinfo {year} {2020}),\ 10.1103/physrevresearch.2.043205}\BibitemShut
  {NoStop}%
\bibitem [{\citenamefont {Arute}\ \emph {et~al.}(2019)\citenamefont {Arute},
  \citenamefont {Arya}, \citenamefont {Babbush}, \citenamefont {Bacon},
  \citenamefont {Bardin}, \citenamefont {Barends}, \citenamefont {Biswas},
  \citenamefont {Boixo}, \citenamefont {Brandao}, \citenamefont {Buell},
  \citenamefont {Burkett}, \citenamefont {Chen}, \citenamefont {Chen},
  \citenamefont {Chiaro}, \citenamefont {Collins}, \citenamefont {Courtney},
  \citenamefont {Dunsworth}, \citenamefont {Farhi}, \citenamefont {Foxen},
  \citenamefont {Fowler}, \citenamefont {Gidney}, \citenamefont {Giustina},
  \citenamefont {Graff}, \citenamefont {Guerin}, \citenamefont {Habegger},
  \citenamefont {Harrigan}, \citenamefont {Hartmann}, \citenamefont {Ho},
  \citenamefont {Hoffmann}, \citenamefont {Huang}, \citenamefont {Humble},
  \citenamefont {Isakov}, \citenamefont {Jeffrey}, \citenamefont {Jiang},
  \citenamefont {Kafri}, \citenamefont {Kechedzhi}, \citenamefont {Kelly},
  \citenamefont {Klimov}, \citenamefont {Knysh}, \citenamefont {Korotkov},
  \citenamefont {Kostritsa}, \citenamefont {Landhuis}, \citenamefont
  {Lindmark}, \citenamefont {Lucero}, \citenamefont {Lyakh}, \citenamefont
  {Mandra}, \citenamefont {McClean}, \citenamefont {McEwen}, \citenamefont
  {Megrant}, \citenamefont {Mi}, \citenamefont {Michielsen}, \citenamefont
  {Mohseni}, \citenamefont {Mutus}, \citenamefont {Naaman}, \citenamefont
  {Neeley}, \citenamefont {Neill}, \citenamefont {Niu}, \citenamefont {Ostby},
  \citenamefont {Petukhov}, \citenamefont {Platt}, \citenamefont {Quintana},
  \citenamefont {Rieffel}, \citenamefont {Roushan}, \citenamefont {Rubin},
  \citenamefont {Sank}, \citenamefont {Satzinger}, \citenamefont {Smelyanskiy},
  \citenamefont {Sung}, \citenamefont {Trevithick}, \citenamefont
  {Vainsencher}, \citenamefont {Villalonga}, \citenamefont {White},
  \citenamefont {Yao}, \citenamefont {Yeh}, \citenamefont {Zalcman},
  \citenamefont {Neven},\ and\ \citenamefont {Martinis}}]{Arute2019}%
  \BibitemOpen
  \bibfield  {author} {\bibinfo {author} {\bibfnamefont {F.}~\bibnamefont
  {Arute}}, \bibinfo {author} {\bibfnamefont {K.}~\bibnamefont {Arya}},
  \bibinfo {author} {\bibfnamefont {R.}~\bibnamefont {Babbush}}, \bibinfo
  {author} {\bibfnamefont {D.}~\bibnamefont {Bacon}}, \bibinfo {author}
  {\bibfnamefont {J.~C.}\ \bibnamefont {Bardin}}, \bibinfo {author}
  {\bibfnamefont {R.}~\bibnamefont {Barends}}, \bibinfo {author} {\bibfnamefont
  {R.}~\bibnamefont {Biswas}}, \bibinfo {author} {\bibfnamefont
  {S.}~\bibnamefont {Boixo}}, \bibinfo {author} {\bibfnamefont {F.~G. S.~L.}\
  \bibnamefont {Brandao}}, \bibinfo {author} {\bibfnamefont {D.~A.}\
  \bibnamefont {Buell}}, \bibinfo {author} {\bibfnamefont {B.}~\bibnamefont
  {Burkett}}, \bibinfo {author} {\bibfnamefont {Y.}~\bibnamefont {Chen}},
  \bibinfo {author} {\bibfnamefont {Z.}~\bibnamefont {Chen}}, \bibinfo {author}
  {\bibfnamefont {B.}~\bibnamefont {Chiaro}}, \bibinfo {author} {\bibfnamefont
  {R.}~\bibnamefont {Collins}}, \bibinfo {author} {\bibfnamefont
  {W.}~\bibnamefont {Courtney}}, \bibinfo {author} {\bibfnamefont
  {A.}~\bibnamefont {Dunsworth}}, \bibinfo {author} {\bibfnamefont
  {E.}~\bibnamefont {Farhi}}, \bibinfo {author} {\bibfnamefont
  {B.}~\bibnamefont {Foxen}}, \bibinfo {author} {\bibfnamefont
  {A.}~\bibnamefont {Fowler}}, \bibinfo {author} {\bibfnamefont
  {C.}~\bibnamefont {Gidney}}, \bibinfo {author} {\bibfnamefont
  {M.}~\bibnamefont {Giustina}}, \bibinfo {author} {\bibfnamefont
  {R.}~\bibnamefont {Graff}}, \bibinfo {author} {\bibfnamefont
  {K.}~\bibnamefont {Guerin}}, \bibinfo {author} {\bibfnamefont
  {S.}~\bibnamefont {Habegger}}, \bibinfo {author} {\bibfnamefont {M.~P.}\
  \bibnamefont {Harrigan}}, \bibinfo {author} {\bibfnamefont {M.~J.}\
  \bibnamefont {Hartmann}}, \bibinfo {author} {\bibfnamefont {A.}~\bibnamefont
  {Ho}}, \bibinfo {author} {\bibfnamefont {M.~R.}\ \bibnamefont {Hoffmann}},
  \bibinfo {author} {\bibfnamefont {T.}~\bibnamefont {Huang}}, \bibinfo
  {author} {\bibfnamefont {T.~S.}\ \bibnamefont {Humble}}, \bibinfo {author}
  {\bibfnamefont {S.~V.}\ \bibnamefont {Isakov}}, \bibinfo {author}
  {\bibfnamefont {E.}~\bibnamefont {Jeffrey}}, \bibinfo {author} {\bibfnamefont
  {Z.}~\bibnamefont {Jiang}}, \bibinfo {author} {\bibfnamefont
  {D.}~\bibnamefont {Kafri}}, \bibinfo {author} {\bibfnamefont
  {K.}~\bibnamefont {Kechedzhi}}, \bibinfo {author} {\bibfnamefont
  {J.}~\bibnamefont {Kelly}}, \bibinfo {author} {\bibfnamefont {P.~V.}\
  \bibnamefont {Klimov}}, \bibinfo {author} {\bibfnamefont {S.}~\bibnamefont
  {Knysh}}, \bibinfo {author} {\bibfnamefont {A.~N.}\ \bibnamefont {Korotkov}},
  \bibinfo {author} {\bibfnamefont {F.}~\bibnamefont {Kostritsa}}, \bibinfo
  {author} {\bibfnamefont {D.}~\bibnamefont {Landhuis}}, \bibinfo {author}
  {\bibfnamefont {M.}~\bibnamefont {Lindmark}}, \bibinfo {author}
  {\bibfnamefont {E.}~\bibnamefont {Lucero}}, \bibinfo {author} {\bibfnamefont
  {D.}~\bibnamefont {Lyakh}}, \bibinfo {author} {\bibfnamefont
  {S.}~\bibnamefont {Mandra}}, \bibinfo {author} {\bibfnamefont {J.~R.}\
  \bibnamefont {McClean}}, \bibinfo {author} {\bibfnamefont {M.}~\bibnamefont
  {McEwen}}, \bibinfo {author} {\bibfnamefont {A.}~\bibnamefont {Megrant}},
  \bibinfo {author} {\bibfnamefont {X.}~\bibnamefont {Mi}}, \bibinfo {author}
  {\bibfnamefont {K.}~\bibnamefont {Michielsen}}, \bibinfo {author}
  {\bibfnamefont {M.}~\bibnamefont {Mohseni}}, \bibinfo {author} {\bibfnamefont
  {J.}~\bibnamefont {Mutus}}, \bibinfo {author} {\bibfnamefont
  {O.}~\bibnamefont {Naaman}}, \bibinfo {author} {\bibfnamefont
  {M.}~\bibnamefont {Neeley}}, \bibinfo {author} {\bibfnamefont
  {C.}~\bibnamefont {Neill}}, \bibinfo {author} {\bibfnamefont {M.~Y.}\
  \bibnamefont {Niu}}, \bibinfo {author} {\bibfnamefont {E.}~\bibnamefont
  {Ostby}}, \bibinfo {author} {\bibfnamefont {A.}~\bibnamefont {Petukhov}},
  \bibinfo {author} {\bibfnamefont {J.~C.}\ \bibnamefont {Platt}}, \bibinfo
  {author} {\bibfnamefont {C.}~\bibnamefont {Quintana}}, \bibinfo {author}
  {\bibfnamefont {E.~G.}\ \bibnamefont {Rieffel}}, \bibinfo {author}
  {\bibfnamefont {P.}~\bibnamefont {Roushan}}, \bibinfo {author} {\bibfnamefont
  {N.~C.}\ \bibnamefont {Rubin}}, \bibinfo {author} {\bibfnamefont
  {D.}~\bibnamefont {Sank}}, \bibinfo {author} {\bibfnamefont {K.~J.}\
  \bibnamefont {Satzinger}}, \bibinfo {author} {\bibfnamefont {V.}~\bibnamefont
  {Smelyanskiy}}, \bibinfo {author} {\bibfnamefont {K.~J.}\ \bibnamefont
  {Sung}}, \bibinfo {author} {\bibfnamefont {M.~D.}\ \bibnamefont
  {Trevithick}}, \bibinfo {author} {\bibfnamefont {A.}~\bibnamefont
  {Vainsencher}}, \bibinfo {author} {\bibfnamefont {B.}~\bibnamefont
  {Villalonga}}, \bibinfo {author} {\bibfnamefont {T.}~\bibnamefont {White}},
  \bibinfo {author} {\bibfnamefont {Z.~J.}\ \bibnamefont {Yao}}, \bibinfo
  {author} {\bibfnamefont {P.}~\bibnamefont {Yeh}}, \bibinfo {author}
  {\bibfnamefont {A.}~\bibnamefont {Zalcman}}, \bibinfo {author} {\bibfnamefont
  {H.}~\bibnamefont {Neven}}, \ and\ \bibinfo {author} {\bibfnamefont {J.~M.}\
  \bibnamefont {Martinis}},\ }\href@noop {} {\bibfield  {journal} {\bibinfo
  {journal} {Nature}\ }\textbf {\bibinfo {volume} {574}} (\bibinfo {year}
  {2019})}\BibitemShut {NoStop}%
\bibitem [{\citenamefont {Asaad}\ \emph {et~al.}(2016)\citenamefont {Asaad},
  \citenamefont {Dickel}, \citenamefont {Langford}, \citenamefont {Poletto},
  \citenamefont {Bruno}, \citenamefont {Rol}, \citenamefont {Deurloo},\ and\
  \citenamefont {DiCarlo}}]{Asaad_2016}%
  \BibitemOpen
  \bibfield  {author} {\bibinfo {author} {\bibfnamefont {S.}~\bibnamefont
  {Asaad}}, \bibinfo {author} {\bibfnamefont {C.}~\bibnamefont {Dickel}},
  \bibinfo {author} {\bibfnamefont {N.~K.}\ \bibnamefont {Langford}}, \bibinfo
  {author} {\bibfnamefont {S.}~\bibnamefont {Poletto}}, \bibinfo {author}
  {\bibfnamefont {A.}~\bibnamefont {Bruno}}, \bibinfo {author} {\bibfnamefont
  {M.~A.}\ \bibnamefont {Rol}}, \bibinfo {author} {\bibfnamefont
  {D.}~\bibnamefont {Deurloo}}, \ and\ \bibinfo {author} {\bibfnamefont
  {L.}~\bibnamefont {DiCarlo}},\ }\href {\doibase 10.1038/npjqi.2016.29}
  {\bibfield  {journal} {\bibinfo  {journal} {npj Quantum Information}\
  }\textbf {\bibinfo {volume} {2}} (\bibinfo {year} {2016}),\
  10.1038/npjqi.2016.29}\BibitemShut {NoStop}%
\bibitem [{\citenamefont {Walter}\ \emph {et~al.}(2017)\citenamefont {Walter},
  \citenamefont {Kurpiers}, \citenamefont {Gasparinetti}, \citenamefont
  {Magnard}, \citenamefont {Potočnik}, \citenamefont {Salathé}, \citenamefont
  {Pechal}, \citenamefont {Mondal}, \citenamefont {Oppliger}, \citenamefont
  {Eichler},\ and\ \citenamefont {et~al.}}]{Walter_2017}%
  \BibitemOpen
  \bibfield  {author} {\bibinfo {author} {\bibfnamefont {T.}~\bibnamefont
  {Walter}}, \bibinfo {author} {\bibfnamefont {P.}~\bibnamefont {Kurpiers}},
  \bibinfo {author} {\bibfnamefont {S.}~\bibnamefont {Gasparinetti}}, \bibinfo
  {author} {\bibfnamefont {P.}~\bibnamefont {Magnard}}, \bibinfo {author}
  {\bibfnamefont {A.}~\bibnamefont {Potočnik}}, \bibinfo {author}
  {\bibfnamefont {Y.}~\bibnamefont {Salathé}}, \bibinfo {author}
  {\bibfnamefont {M.}~\bibnamefont {Pechal}}, \bibinfo {author} {\bibfnamefont
  {M.}~\bibnamefont {Mondal}}, \bibinfo {author} {\bibfnamefont
  {M.}~\bibnamefont {Oppliger}}, \bibinfo {author} {\bibfnamefont
  {C.}~\bibnamefont {Eichler}}, \ and\ \bibinfo {author} {\bibnamefont
  {et~al.}},\ }\href {\doibase 10.1103/physrevapplied.7.054020} {\bibfield
  {journal} {\bibinfo  {journal} {Physical Review Applied}\ }\textbf {\bibinfo
  {volume} {7}} (\bibinfo {year} {2017}),\
  10.1103/physrevapplied.7.054020}\BibitemShut {NoStop}%
\bibitem [{\citenamefont {Ofek}\ \emph {et~al.}(2016)\citenamefont {Ofek},
  \citenamefont {Petrenko}, \citenamefont {Heeres}, \citenamefont {Reinhold},
  \citenamefont {Leghtas}, \citenamefont {Vlastakis}, \citenamefont {Liu},
  \citenamefont {Frunzio}, \citenamefont {Girvin}, \citenamefont {Jiang},
  \citenamefont {Mirrahimi}, \citenamefont {Devoret},\ and\ \citenamefont
  {Schoelkopf}}]{Ofek2016}%
  \BibitemOpen
  \bibfield  {author} {\bibinfo {author} {\bibfnamefont {N.}~\bibnamefont
  {Ofek}}, \bibinfo {author} {\bibfnamefont {A.}~\bibnamefont {Petrenko}},
  \bibinfo {author} {\bibfnamefont {R.}~\bibnamefont {Heeres}}, \bibinfo
  {author} {\bibfnamefont {P.}~\bibnamefont {Reinhold}}, \bibinfo {author}
  {\bibfnamefont {Z.}~\bibnamefont {Leghtas}}, \bibinfo {author} {\bibfnamefont
  {B.}~\bibnamefont {Vlastakis}}, \bibinfo {author} {\bibfnamefont
  {Y.}~\bibnamefont {Liu}}, \bibinfo {author} {\bibfnamefont {L.}~\bibnamefont
  {Frunzio}}, \bibinfo {author} {\bibfnamefont {S.~M.}\ \bibnamefont {Girvin}},
  \bibinfo {author} {\bibfnamefont {L.}~\bibnamefont {Jiang}}, \bibinfo
  {author} {\bibfnamefont {M.}~\bibnamefont {Mirrahimi}}, \bibinfo {author}
  {\bibfnamefont {M.~H.}\ \bibnamefont {Devoret}}, \ and\ \bibinfo {author}
  {\bibfnamefont {R.~J.}\ \bibnamefont {Schoelkopf}},\ }\href {\doibase
  10.1038/nature18949} {\bibfield  {journal} {\bibinfo  {journal} {Nature}\
  }\textbf {\bibinfo {volume} {536}},\ \bibinfo {pages} {441} (\bibinfo {year}
  {2016})}\BibitemShut {NoStop}%
\bibitem [{art(2021)}]{artiq}%
  \BibitemOpen
  \href@noop {} {\enquote {\bibinfo {title} {{ARTIQ} {W}ebsite},}\ }\bibinfo
  {howpublished} {\url{https://m-labs.hk/experiment-control/artiq/ }} (\bibinfo
  {year} {2021})\BibitemShut {NoStop}%
\bibitem [{\citenamefont {Ryan}\ \emph {et~al.}(2017)\citenamefont {Ryan},
  \citenamefont {Johnson}, \citenamefont {Rist{\`{e}}}, \citenamefont
  {Donovan},\ and\ \citenamefont {Ohki}}]{Ryan2017}%
  \BibitemOpen
  \bibfield  {author} {\bibinfo {author} {\bibfnamefont {C.~A.}\ \bibnamefont
  {Ryan}}, \bibinfo {author} {\bibfnamefont {B.~R.}\ \bibnamefont {Johnson}},
  \bibinfo {author} {\bibfnamefont {D.}~\bibnamefont {Rist{\`{e}}}}, \bibinfo
  {author} {\bibfnamefont {B.}~\bibnamefont {Donovan}}, \ and\ \bibinfo
  {author} {\bibfnamefont {T.~A.}\ \bibnamefont {Ohki}},\ }\href {\doibase
  10.1063/1.5006525} {\bibfield  {journal} {\bibinfo  {journal} {Review of
  Scientific Instruments}\ }\textbf {\bibinfo {volume} {88}} (\bibinfo {year}
  {2017}),\ 10.1063/1.5006525}\BibitemShut {NoStop}%
\bibitem [{key(2021)}]{keysight}%
  \BibitemOpen
  \href@noop {} {\enquote {\bibinfo {title} {Keysight {Q}uantum {S}olutions},}\
  }\bibinfo {howpublished}
  {\url{https://www.keysight.com/us/en/solutions/emerging-technologies/quantum-solutions.html}}
  (\bibinfo {year} {2021})\BibitemShut {NoStop}%
\bibitem [{zur(2021)}]{zurichinst}%
  \BibitemOpen
  \href@noop {} {\enquote {\bibinfo {title} {{Z}urich {I}nstruments
  {W}ebsite},}\ }\bibinfo {howpublished}
  {\url{https://www.zhinst.com/americas/en}} (\bibinfo {year}
  {2021})\BibitemShut {NoStop}%
\bibitem [{qua(2021)}]{quantummachines}%
  \BibitemOpen
  \href@noop {} {\enquote {\bibinfo {title} {{Q}uantum {M}achines {W}ebsite},}\
  }\bibinfo {howpublished} {\url{https://www.quantum-machines.co/}} (\bibinfo
  {year} {2021})\BibitemShut {NoStop}%
\bibitem [{\citenamefont {Xu}\ \emph {et~al.}(2021)\citenamefont {Xu},
  \citenamefont {Huang}, \citenamefont {Balewski}, \citenamefont {Naik},
  \citenamefont {Morvan}, \citenamefont {Mitchell}, \citenamefont {Nowrouzi},
  \citenamefont {Santiago},\ and\ \citenamefont {Siddiqi}}]{xu2021}%
  \BibitemOpen
  \bibfield  {author} {\bibinfo {author} {\bibfnamefont {Y.}~\bibnamefont
  {Xu}}, \bibinfo {author} {\bibfnamefont {G.}~\bibnamefont {Huang}}, \bibinfo
  {author} {\bibfnamefont {J.}~\bibnamefont {Balewski}}, \bibinfo {author}
  {\bibfnamefont {R.}~\bibnamefont {Naik}}, \bibinfo {author} {\bibfnamefont
  {A.}~\bibnamefont {Morvan}}, \bibinfo {author} {\bibfnamefont
  {B.}~\bibnamefont {Mitchell}}, \bibinfo {author} {\bibfnamefont
  {K.}~\bibnamefont {Nowrouzi}}, \bibinfo {author} {\bibfnamefont {D.~I.}\
  \bibnamefont {Santiago}}, \ and\ \bibinfo {author} {\bibfnamefont
  {I.}~\bibnamefont {Siddiqi}},\ }\href@noop {} {\enquote {\bibinfo {title}
  {Qubic: An open source fpga-based control and measurement system for
  superconducting quantum information processors},}\ } (\bibinfo {year}
  {2021}),\ \Eprint {http://arxiv.org/abs/2101.00071} {arXiv:2101.00071
  [quant-ph]} \BibitemShut {NoStop}%
\bibitem [{\citenamefont {Kalfus}\ \emph {et~al.}(2021)\citenamefont {Kalfus},
  \citenamefont {Lee}, \citenamefont {Ribeill}, \citenamefont {Fallek},
  \citenamefont {Wagner}, \citenamefont {Donovan}, \citenamefont {Riste},\ and\
  \citenamefont {Ohki}}]{Kalfus2021}%
  \BibitemOpen
  \bibfield  {author} {\bibinfo {author} {\bibfnamefont {W.~D.}\ \bibnamefont
  {Kalfus}}, \bibinfo {author} {\bibfnamefont {D.~F.}\ \bibnamefont {Lee}},
  \bibinfo {author} {\bibfnamefont {G.~J.}\ \bibnamefont {Ribeill}}, \bibinfo
  {author} {\bibfnamefont {S.~D.}\ \bibnamefont {Fallek}}, \bibinfo {author}
  {\bibfnamefont {A.}~\bibnamefont {Wagner}}, \bibinfo {author} {\bibfnamefont
  {B.}~\bibnamefont {Donovan}}, \bibinfo {author} {\bibfnamefont
  {D.}~\bibnamefont {Riste}}, \ and\ \bibinfo {author} {\bibfnamefont {T.~A.}\
  \bibnamefont {Ohki}},\ }\href {\doibase 10.1109/tqe.2020.3042895} {\bibfield
  {journal} {\bibinfo  {journal} {IEEE Transactions on Quantum Engineering}\
  }\textbf {\bibinfo {volume} {1}} (\bibinfo {year} {2021}),\
  10.1109/tqe.2020.3042895}\BibitemShut {NoStop}%
\bibitem [{\citenamefont {Jolin}\ \emph {et~al.}(2020)\citenamefont {Jolin},
  \citenamefont {Borgani}, \citenamefont {Tholén}, \citenamefont
  {Forchheimer},\ and\ \citenamefont {Haviland}}]{Jolin2020}%
  \BibitemOpen
  \bibfield  {author} {\bibinfo {author} {\bibfnamefont {S.~W.}\ \bibnamefont
  {Jolin}}, \bibinfo {author} {\bibfnamefont {R.}~\bibnamefont {Borgani}},
  \bibinfo {author} {\bibfnamefont {M.~O.}\ \bibnamefont {Tholén}}, \bibinfo
  {author} {\bibfnamefont {D.}~\bibnamefont {Forchheimer}}, \ and\ \bibinfo
  {author} {\bibfnamefont {D.~B.}\ \bibnamefont {Haviland}},\ }\href {\doibase
  10.1063/5.0025836} {\bibfield  {journal} {\bibinfo  {journal} {Review of
  Scientific Instruments}\ }\textbf {\bibinfo {volume} {91}},\ \bibinfo {pages}
  {124707} (\bibinfo {year} {2020})}\BibitemShut {NoStop}%
\bibitem [{rfs(2021)}]{rfsocwebsite}%
  \BibitemOpen
  \href@noop {} {\enquote {\bibinfo {title} {{X}ilinx {RFSOC} {W}ebsite},}\
  }\bibinfo {howpublished}
  {\url{https://www.xilinx.com/products/silicon-devices/soc/rfsoc.html}}
  (\bibinfo {year} {2021})\BibitemShut {NoStop}%
\bibitem [{\citenamefont {Gebauer}\ \emph {et~al.}(2021)\citenamefont
  {Gebauer}, \citenamefont {Karcher},\ and\ \citenamefont
  {Sander}}]{Gebauer2021}%
  \BibitemOpen
  \bibfield  {author} {\bibinfo {author} {\bibfnamefont {R.}~\bibnamefont
  {Gebauer}}, \bibinfo {author} {\bibfnamefont {N.}~\bibnamefont {Karcher}}, \
  and\ \bibinfo {author} {\bibfnamefont {O.}~\bibnamefont {Sander}},\ }in\
  \href {\doibase 10.1109/ICFPT52863.2021.9609909} {\emph {\bibinfo {booktitle}
  {2021 International Conference on Field-Programmable Technology (ICFPT)}}}\
  (\bibinfo {year} {2021})\ pp.\ \bibinfo {pages} {1--9}\BibitemShut {NoStop}%
\bibitem [{\citenamefont {Park}\ \emph {et~al.}(2021)\citenamefont {Park},
  \citenamefont {Yap}, \citenamefont {Tan}, \citenamefont {Hufnagel},
  \citenamefont {Nguyen}, \citenamefont {Lau}, \citenamefont {Efthymiou},
  \citenamefont {Carrazza}, \citenamefont {Budoyo},\ and\ \citenamefont
  {Dumke}}]{park2021icarusq}%
  \BibitemOpen
  \bibfield  {author} {\bibinfo {author} {\bibfnamefont {K.~H.}\ \bibnamefont
  {Park}}, \bibinfo {author} {\bibfnamefont {Y.~S.}\ \bibnamefont {Yap}},
  \bibinfo {author} {\bibfnamefont {Y.~P.}\ \bibnamefont {Tan}}, \bibinfo
  {author} {\bibfnamefont {C.}~\bibnamefont {Hufnagel}}, \bibinfo {author}
  {\bibfnamefont {L.~H.}\ \bibnamefont {Nguyen}}, \bibinfo {author}
  {\bibfnamefont {K.~H.}\ \bibnamefont {Lau}}, \bibinfo {author} {\bibfnamefont
  {S.}~\bibnamefont {Efthymiou}}, \bibinfo {author} {\bibfnamefont
  {S.}~\bibnamefont {Carrazza}}, \bibinfo {author} {\bibfnamefont {R.~P.}\
  \bibnamefont {Budoyo}}, \ and\ \bibinfo {author} {\bibfnamefont
  {R.}~\bibnamefont {Dumke}},\ }\href@noop {} {\enquote {\bibinfo {title}
  {Icarus-q: A scalable rfsoc-based control system for superconducting quantum
  computers},}\ } (\bibinfo {year} {2021}),\ \Eprint
  {http://arxiv.org/abs/2112.02933} {arXiv:2112.02933 [quant-ph]} \BibitemShut
  {NoStop}%
\bibitem [{imp(2021)}]{imp}%
  \BibitemOpen
  \href@noop {} {\enquote {\bibinfo {title} {{I}ntermodulation {P}roducts
  {W}ebsite},}\ }\bibinfo {howpublished}
  {\url{https://intermodulation-products.com/products/microwave-platforms }}
  (\bibinfo {year} {2021})\BibitemShut {NoStop}%
\bibitem [{pyn(2021)}]{pynqwebsite}%
  \BibitemOpen
  \href@noop {} {\enquote {\bibinfo {title} {{P}ython productivitye for
  z{ynq}},}\ }\bibinfo {howpublished} {\url{https://pynq.io}} (\bibinfo {year}
  {2021})\BibitemShut {NoStop}%
\bibitem [{rfs(2020)}]{rfsocdatasheet}%
  \BibitemOpen
  \href@noop {} {\enquote {\bibinfo {title} {Zynq {U}ltrascale+ {RFSOC}
  {P}roduct {D}ata {S}heet: {O}verview ({DS}889)},}\ }\bibinfo {howpublished}
  {\url{https://www.xilinx.com/support/documentation/data_sheets/ds889-zynq-usp-rfsoc-overview.pdf}}
  (\bibinfo {year} {2020})\BibitemShut {NoStop}%
\bibitem [{QIC(2021{\natexlab{a}})}]{QICKrepo}%
  \BibitemOpen
  \href@noop {} {\enquote {\bibinfo {title} {{QICK} {G}ithub repository},}\
  }\bibinfo {howpublished} {\url{https://github.com/openquantumhardware/qick}}
  (\bibinfo {year} {2021}{\natexlab{a}})\BibitemShut {NoStop}%
\bibitem [{\citenamefont {Dixit}\ \emph {et~al.}(2021)\citenamefont {Dixit},
  \citenamefont {Chakram}, \citenamefont {He}, \citenamefont {Agrawal},
  \citenamefont {Naik}, \citenamefont {Schuster},\ and\ \citenamefont
  {Chou}}]{Dixit2021}%
  \BibitemOpen
  \bibfield  {author} {\bibinfo {author} {\bibfnamefont {A.~V.}\ \bibnamefont
  {Dixit}}, \bibinfo {author} {\bibfnamefont {S.}~\bibnamefont {Chakram}},
  \bibinfo {author} {\bibfnamefont {K.}~\bibnamefont {He}}, \bibinfo {author}
  {\bibfnamefont {A.}~\bibnamefont {Agrawal}}, \bibinfo {author} {\bibfnamefont
  {R.~K.}\ \bibnamefont {Naik}}, \bibinfo {author} {\bibfnamefont {D.~I.}\
  \bibnamefont {Schuster}}, \ and\ \bibinfo {author} {\bibfnamefont
  {A.}~\bibnamefont {Chou}},\ }\href {\doibase 10.1103/PhysRevLett.126.141302}
  {\bibfield  {journal} {\bibinfo  {journal} {Physical Review Letters}\
  }\textbf {\bibinfo {volume} {126}} (\bibinfo {year} {2021}),\
  10.1103/PhysRevLett.126.141302}\BibitemShut {NoStop}%
\bibitem [{\citenamefont {Walter}\ \emph {et~al.}(2020)\citenamefont {Walter},
  \citenamefont {Fruitwala}, \citenamefont {Steiger}, \citenamefont {Bailey},
  \citenamefont {Zobrist}, \citenamefont {Swimmer}, \citenamefont {Lipartito},
  \citenamefont {Smith}, \citenamefont {Meeker}, \citenamefont {Bockstiegel},
  \citenamefont {Coiffard}, \citenamefont {Dodkins}, \citenamefont {Szypryt},
  \citenamefont {Davis}, \citenamefont {Daal}, \citenamefont {Bumble},
  \citenamefont {Collura}, \citenamefont {Guyon}, \citenamefont {Lozi},
  \citenamefont {Vievard}, \citenamefont {Jovanovic}, \citenamefont
  {Martinache}, \citenamefont {Currie},\ and\ \citenamefont
  {Mazin}}]{Walter2020}%
  \BibitemOpen
  \bibfield  {author} {\bibinfo {author} {\bibfnamefont {A.~B.}\ \bibnamefont
  {Walter}}, \bibinfo {author} {\bibfnamefont {N.}~\bibnamefont {Fruitwala}},
  \bibinfo {author} {\bibfnamefont {S.}~\bibnamefont {Steiger}}, \bibinfo
  {author} {\bibfnamefont {J.~I.}\ \bibnamefont {Bailey}}, \bibinfo {author}
  {\bibfnamefont {N.}~\bibnamefont {Zobrist}}, \bibinfo {author} {\bibfnamefont
  {N.}~\bibnamefont {Swimmer}}, \bibinfo {author} {\bibfnamefont
  {I.}~\bibnamefont {Lipartito}}, \bibinfo {author} {\bibfnamefont {J.~P.}\
  \bibnamefont {Smith}}, \bibinfo {author} {\bibfnamefont {S.~R.}\ \bibnamefont
  {Meeker}}, \bibinfo {author} {\bibfnamefont {C.}~\bibnamefont {Bockstiegel}},
  \bibinfo {author} {\bibfnamefont {G.}~\bibnamefont {Coiffard}}, \bibinfo
  {author} {\bibfnamefont {R.}~\bibnamefont {Dodkins}}, \bibinfo {author}
  {\bibfnamefont {P.}~\bibnamefont {Szypryt}}, \bibinfo {author} {\bibfnamefont
  {K.~K.}\ \bibnamefont {Davis}}, \bibinfo {author} {\bibfnamefont
  {M.}~\bibnamefont {Daal}}, \bibinfo {author} {\bibfnamefont {B.}~\bibnamefont
  {Bumble}}, \bibinfo {author} {\bibfnamefont {G.}~\bibnamefont {Collura}},
  \bibinfo {author} {\bibfnamefont {O.}~\bibnamefont {Guyon}}, \bibinfo
  {author} {\bibfnamefont {J.}~\bibnamefont {Lozi}}, \bibinfo {author}
  {\bibfnamefont {S.}~\bibnamefont {Vievard}}, \bibinfo {author} {\bibfnamefont
  {N.}~\bibnamefont {Jovanovic}}, \bibinfo {author} {\bibfnamefont
  {F.}~\bibnamefont {Martinache}}, \bibinfo {author} {\bibfnamefont
  {T.}~\bibnamefont {Currie}}, \ and\ \bibinfo {author} {\bibfnamefont {B.~A.}\
  \bibnamefont {Mazin}},\ }\href {\doibase 10.1088/1538-3873/abc60f} {\bibfield
   {journal} {\bibinfo  {journal} {Publications of the Astronomical Society of
  the Pacific}\ }\textbf {\bibinfo {volume} {132}} (\bibinfo {year} {2020}),\
  10.1088/1538-3873/abc60f}\BibitemShut {NoStop}%
\bibitem [{\citenamefont {Fruitwala}\ \emph {et~al.}(2020)\citenamefont
  {Fruitwala}, \citenamefont {Strader}, \citenamefont {Cancelo}, \citenamefont
  {Zmuda}, \citenamefont {Treptow}, \citenamefont {Wilcer}, \citenamefont
  {Stoughton}, \citenamefont {Walter}, \citenamefont {Zobrist}, \citenamefont
  {Collura}, \citenamefont {Lipartito}, \citenamefont {Bailey},\ and\
  \citenamefont {Mazin}}]{Fruitwala2020}%
  \BibitemOpen
  \bibfield  {author} {\bibinfo {author} {\bibfnamefont {N.}~\bibnamefont
  {Fruitwala}}, \bibinfo {author} {\bibfnamefont {P.}~\bibnamefont {Strader}},
  \bibinfo {author} {\bibfnamefont {G.}~\bibnamefont {Cancelo}}, \bibinfo
  {author} {\bibfnamefont {T.}~\bibnamefont {Zmuda}}, \bibinfo {author}
  {\bibfnamefont {K.}~\bibnamefont {Treptow}}, \bibinfo {author} {\bibfnamefont
  {N.}~\bibnamefont {Wilcer}}, \bibinfo {author} {\bibfnamefont
  {C.}~\bibnamefont {Stoughton}}, \bibinfo {author} {\bibfnamefont {A.~B.}\
  \bibnamefont {Walter}}, \bibinfo {author} {\bibfnamefont {N.}~\bibnamefont
  {Zobrist}}, \bibinfo {author} {\bibfnamefont {G.}~\bibnamefont {Collura}},
  \bibinfo {author} {\bibfnamefont {I.}~\bibnamefont {Lipartito}}, \bibinfo
  {author} {\bibfnamefont {J.~I.}\ \bibnamefont {Bailey}}, \ and\ \bibinfo
  {author} {\bibfnamefont {B.~A.}\ \bibnamefont {Mazin}},\ }\href {\doibase
  10.1063/5.0029457} {\bibfield  {journal} {\bibinfo  {journal} {Review of
  Scientific Instruments}\ }\textbf {\bibinfo {volume} {91}} (\bibinfo {year}
  {2020}),\ 10.1063/5.0029457}\BibitemShut {NoStop}%
\bibitem [{\citenamefont {Zhang}\ \emph {et~al.}(2021)\citenamefont {Zhang},
  \citenamefont {Chakram}, \citenamefont {Roy}, \citenamefont {Earnest},
  \citenamefont {Lu}, \citenamefont {Huang}, \citenamefont {Weiss},
  \citenamefont {Koch},\ and\ \citenamefont {Schuster}}]{Zhang2021}%
  \BibitemOpen
  \bibfield  {author} {\bibinfo {author} {\bibfnamefont {H.}~\bibnamefont
  {Zhang}}, \bibinfo {author} {\bibfnamefont {S.}~\bibnamefont {Chakram}},
  \bibinfo {author} {\bibfnamefont {T.}~\bibnamefont {Roy}}, \bibinfo {author}
  {\bibfnamefont {N.}~\bibnamefont {Earnest}}, \bibinfo {author} {\bibfnamefont
  {Y.}~\bibnamefont {Lu}}, \bibinfo {author} {\bibfnamefont {Z.}~\bibnamefont
  {Huang}}, \bibinfo {author} {\bibfnamefont {D.~K.}\ \bibnamefont {Weiss}},
  \bibinfo {author} {\bibfnamefont {J.}~\bibnamefont {Koch}}, \ and\ \bibinfo
  {author} {\bibfnamefont {D.~I.}\ \bibnamefont {Schuster}},\ }\href {\doibase
  10.1103/PhysRevX.11.011010} {\bibfield  {journal} {\bibinfo  {journal}
  {Physical Review X}\ }\textbf {\bibinfo {volume} {11}} (\bibinfo {year}
  {2021}),\ 10.1103/PhysRevX.11.011010}\BibitemShut {NoStop}%
\bibitem [{\citenamefont {Hanson}\ \emph {et~al.}(2007)\citenamefont {Hanson},
  \citenamefont {Petta}, \citenamefont {Tarucha},\ and\ \citenamefont
  {Vandersypen}}]{Hanson2007}%
  \BibitemOpen
  \bibfield  {author} {\bibinfo {author} {\bibfnamefont {R.}~\bibnamefont
  {Hanson}}, \bibinfo {author} {\bibfnamefont {J.~R.}\ \bibnamefont {Petta}},
  \bibinfo {author} {\bibfnamefont {S.}~\bibnamefont {Tarucha}}, \ and\
  \bibinfo {author} {\bibfnamefont {L.~M.~K.}\ \bibnamefont {Vandersypen}},\
  }\href {\doibase 10.1103/revmodphys.79.1217} {\bibfield  {journal} {\bibinfo
  {journal} {Reviews of Modern Physics}\ }\textbf {\bibinfo {volume} {79}},\
  \bibinfo {pages} {1217} (\bibinfo {year} {2007})}\BibitemShut {NoStop}%
\bibitem [{QIC(2021{\natexlab{b}})}]{QICKdoc}%
  \BibitemOpen
  \href@noop {} {\enquote {\bibinfo {title} {{QICK} documentation website},}\
  }\bibinfo {howpublished} {\url{https://qick-docs.readthedocs.io/en/latest/}}
  (\bibinfo {year} {2021}{\natexlab{b}})\BibitemShut {NoStop}%
\bibitem [{qis(2021)}]{qiskit}%
  \BibitemOpen
  \href@noop {} {\enquote {\bibinfo {title} {{IBM} {Q}iskit library},}\
  }\bibinfo {howpublished} {\url{https://qiskit.org/}} (\bibinfo {year}
  {2021})\BibitemShut {NoStop}%
\bibitem [{SHF(2021)}]{SHFQA}%
  \BibitemOpen
  \href@noop {} {\enquote {\bibinfo {title} {{Z}urich {I}nstruments {SHFQA}
  {S}pecifications},}\ }\bibinfo {howpublished}
  {\url{https://docs.zhinst.com/shfqa_user_manual/specifications.html }}
  (\bibinfo {year} {2021})\BibitemShut {NoStop}%
\bibitem [{\citenamefont {Dijk}\ \emph {et~al.}(2020)\citenamefont {Dijk},
  \citenamefont {Patra}, \citenamefont {Pellerano}, \citenamefont {Charbon},
  \citenamefont {Sebastiano},\ and\ \citenamefont {Babaie}}]{VanDijk2020}%
  \BibitemOpen
  \bibfield  {author} {\bibinfo {author} {\bibfnamefont {J.~P.~V.}\
  \bibnamefont {Dijk}}, \bibinfo {author} {\bibfnamefont {B.}~\bibnamefont
  {Patra}}, \bibinfo {author} {\bibfnamefont {S.}~\bibnamefont {Pellerano}},
  \bibinfo {author} {\bibfnamefont {E.}~\bibnamefont {Charbon}}, \bibinfo
  {author} {\bibfnamefont {F.}~\bibnamefont {Sebastiano}}, \ and\ \bibinfo
  {author} {\bibfnamefont {M.}~\bibnamefont {Babaie}},\ }\href {\doibase
  10.1109/TCSI.2020.3019413} {\bibfield  {journal} {\bibinfo  {journal} {IEEE
  Transactions on Circuits and Systems I: Regular Papers}\ }\textbf {\bibinfo
  {volume} {67}} (\bibinfo {year} {2020}),\
  10.1109/TCSI.2020.3019413}\BibitemShut {NoStop}%
\bibitem [{\citenamefont {Ball}\ \emph {et~al.}(2016)\citenamefont {Ball},
  \citenamefont {Oliver},\ and\ \citenamefont {Biercuk}}]{Ball2016}%
  \BibitemOpen
  \bibfield  {author} {\bibinfo {author} {\bibfnamefont {H.}~\bibnamefont
  {Ball}}, \bibinfo {author} {\bibfnamefont {W.~D.}\ \bibnamefont {Oliver}}, \
  and\ \bibinfo {author} {\bibfnamefont {M.~J.}\ \bibnamefont {Biercuk}},\
  }\href {\doibase 10.1038/npjqi.2016.33} {\enquote {\bibinfo {title} {The role
  of master clock stability in quantum information processing},}\ } (\bibinfo
  {year} {2016})\BibitemShut {NoStop}%
\bibitem [{PG2(2021)}]{PG269}%
  \BibitemOpen
  \href@noop {} {\enquote {\bibinfo {title} {Zynq {U}ltrascale+ {RFSOC} {RF}
  {D}ata {C}onverter {P}roduct {G}uide ({PG}269)},}\ }\bibinfo {howpublished}
  {\url{https://www.xilinx.com/support/documentation/ip_documentation/usp_rf_data_converter/v2_6/pg269-rf-data-converter.pdf}}
  (\bibinfo {year} {2021})\BibitemShut {NoStop}%
\bibitem [{\citenamefont {Koch}\ \emph {et~al.}(2007)\citenamefont {Koch},
  \citenamefont {Yu}, \citenamefont {Gambetta}, \citenamefont {Houck},
  \citenamefont {Schuster}, \citenamefont {Majer}, \citenamefont {Blais},
  \citenamefont {Devoret}, \citenamefont {Girvin},\ and\ \citenamefont
  {Schoelkopf}}]{Koch2007}%
  \BibitemOpen
  \bibfield  {author} {\bibinfo {author} {\bibfnamefont {J.}~\bibnamefont
  {Koch}}, \bibinfo {author} {\bibfnamefont {T.~M.}\ \bibnamefont {Yu}},
  \bibinfo {author} {\bibfnamefont {J.}~\bibnamefont {Gambetta}}, \bibinfo
  {author} {\bibfnamefont {A.~A.}\ \bibnamefont {Houck}}, \bibinfo {author}
  {\bibfnamefont {D.~I.}\ \bibnamefont {Schuster}}, \bibinfo {author}
  {\bibfnamefont {J.}~\bibnamefont {Majer}}, \bibinfo {author} {\bibfnamefont
  {A.}~\bibnamefont {Blais}}, \bibinfo {author} {\bibfnamefont {M.~H.}\
  \bibnamefont {Devoret}}, \bibinfo {author} {\bibfnamefont {S.~M.}\
  \bibnamefont {Girvin}}, \ and\ \bibinfo {author} {\bibfnamefont {R.~J.}\
  \bibnamefont {Schoelkopf}},\ }\href {\doibase 10.1103/PhysRevA.76.042319}
  {\bibfield  {journal} {\bibinfo  {journal} {Physical Review A - Atomic,
  Molecular, and Optical Physics}\ }\textbf {\bibinfo {volume} {76}} (\bibinfo
  {year} {2007}),\ 10.1103/PhysRevA.76.042319}\BibitemShut {NoStop}%
\bibitem [{\citenamefont {Gambetta}\ \emph {et~al.}(2006)\citenamefont
  {Gambetta}, \citenamefont {Blais}, \citenamefont {Schuster}, \citenamefont
  {Wallraff}, \citenamefont {Frunzio}, \citenamefont {Majer}, \citenamefont
  {Devoret}, \citenamefont {Girvin},\ and\ \citenamefont
  {Schoelkopf}}]{Gambetta2006}%
  \BibitemOpen
  \bibfield  {author} {\bibinfo {author} {\bibfnamefont {J.}~\bibnamefont
  {Gambetta}}, \bibinfo {author} {\bibfnamefont {A.}~\bibnamefont {Blais}},
  \bibinfo {author} {\bibfnamefont {D.~I.}\ \bibnamefont {Schuster}}, \bibinfo
  {author} {\bibfnamefont {A.}~\bibnamefont {Wallraff}}, \bibinfo {author}
  {\bibfnamefont {L.}~\bibnamefont {Frunzio}}, \bibinfo {author} {\bibfnamefont
  {J.}~\bibnamefont {Majer}}, \bibinfo {author} {\bibfnamefont {M.~H.}\
  \bibnamefont {Devoret}}, \bibinfo {author} {\bibfnamefont {S.~M.}\
  \bibnamefont {Girvin}}, \ and\ \bibinfo {author} {\bibfnamefont {R.~J.}\
  \bibnamefont {Schoelkopf}},\ }\href {\doibase 10.1103/PhysRevA.74.042318}
  {\bibfield  {journal} {\bibinfo  {journal} {Physical Review A - Atomic,
  Molecular, and Optical Physics}\ }\textbf {\bibinfo {volume} {74}} (\bibinfo
  {year} {2006}),\ 10.1103/PhysRevA.74.042318}\BibitemShut {NoStop}%
\bibitem [{\citenamefont {Nakamura}\ \emph {et~al.}(1999)\citenamefont
  {Nakamura}, \citenamefont {Pashkin},\ and\ \citenamefont
  {Tsai}}]{Nakamura1999}%
  \BibitemOpen
  \bibfield  {author} {\bibinfo {author} {\bibfnamefont {Y.}~\bibnamefont
  {Nakamura}}, \bibinfo {author} {\bibfnamefont {Y.~A.}\ \bibnamefont
  {Pashkin}}, \ and\ \bibinfo {author} {\bibfnamefont {J.~S.}\ \bibnamefont
  {Tsai}},\ }\href {\doibase 10.1038/19718} {\bibfield  {journal} {\bibinfo
  {journal} {Nature}\ }\textbf {\bibinfo {volume} {398}} (\bibinfo {year}
  {1999}),\ 10.1038/19718}\BibitemShut {NoStop}%
\bibitem [{\citenamefont {Fowler}\ \emph {et~al.}(2012)\citenamefont {Fowler},
  \citenamefont {Mariantoni}, \citenamefont {Martinis},\ and\ \citenamefont
  {Cleland}}]{Fowler2012}%
  \BibitemOpen
  \bibfield  {author} {\bibinfo {author} {\bibfnamefont {A.~G.}\ \bibnamefont
  {Fowler}}, \bibinfo {author} {\bibfnamefont {M.}~\bibnamefont {Mariantoni}},
  \bibinfo {author} {\bibfnamefont {J.~M.}\ \bibnamefont {Martinis}}, \ and\
  \bibinfo {author} {\bibfnamefont {A.~N.}\ \bibnamefont {Cleland}},\ }\href
  {\doibase 10.1103/PhysRevA.86.032324} {\bibfield  {journal} {\bibinfo
  {journal} {Physical Review A - Atomic, Molecular, and Optical Physics}\
  }\textbf {\bibinfo {volume} {86}} (\bibinfo {year} {2012}),\
  10.1103/PhysRevA.86.032324}\BibitemShut {NoStop}%
\bibitem [{\citenamefont {Wallraff}\ \emph {et~al.}(2005)\citenamefont
  {Wallraff}, \citenamefont {Schuster}, \citenamefont {Blais}, \citenamefont
  {Frunzio}, \citenamefont {Majer}, \citenamefont {Devoret}, \citenamefont
  {Girvin},\ and\ \citenamefont {Schoelkopf}}]{Wallraff2005}%
  \BibitemOpen
  \bibfield  {author} {\bibinfo {author} {\bibfnamefont {A.}~\bibnamefont
  {Wallraff}}, \bibinfo {author} {\bibfnamefont {D.~I.}\ \bibnamefont
  {Schuster}}, \bibinfo {author} {\bibfnamefont {A.}~\bibnamefont {Blais}},
  \bibinfo {author} {\bibfnamefont {L.}~\bibnamefont {Frunzio}}, \bibinfo
  {author} {\bibfnamefont {J.}~\bibnamefont {Majer}}, \bibinfo {author}
  {\bibfnamefont {M.~H.}\ \bibnamefont {Devoret}}, \bibinfo {author}
  {\bibfnamefont {S.~M.}\ \bibnamefont {Girvin}}, \ and\ \bibinfo {author}
  {\bibfnamefont {R.~J.}\ \bibnamefont {Schoelkopf}},\ }\href {\doibase
  10.1103/PhysRevLett.95.060501} {\bibfield  {journal} {\bibinfo  {journal}
  {Physical Review Letters}\ }\textbf {\bibinfo {volume} {95}} (\bibinfo {year}
  {2005}),\ 10.1103/PhysRevLett.95.060501}\BibitemShut {NoStop}%
\bibitem [{\citenamefont {Martinis}(2009)}]{Martinis2009}%
  \BibitemOpen
  \bibfield  {author} {\bibinfo {author} {\bibfnamefont {J.~M.}\ \bibnamefont
  {Martinis}},\ }\href {\doibase 10.1007/s11128-009-0105-1} {\bibfield
  {journal} {\bibinfo  {journal} {Quantum Information Processing}\ }\textbf
  {\bibinfo {volume} {8}} (\bibinfo {year} {2009}),\
  10.1007/s11128-009-0105-1}\BibitemShut {NoStop}%
\bibitem [{\citenamefont {Preskill}(1997)}]{Preskill1997}%
  \BibitemOpen
  \bibfield  {author} {\bibinfo {author} {\bibfnamefont {J.}~\bibnamefont
  {Preskill}},\ }\href@noop {} {\bibfield  {journal} {\bibinfo  {journal}
  {Annual Symposium on Foundations of Computer Science - Proceedings}\ }
  (\bibinfo {year} {1997})}\BibitemShut {NoStop}%
\bibitem [{\citenamefont {DiVincenzo}(2000)}]{DiVincenzo2000}%
  \BibitemOpen
  \bibfield  {author} {\bibinfo {author} {\bibfnamefont {D.~P.}\ \bibnamefont
  {DiVincenzo}},\ }\href {\doibase
  10.1002/1521-3978(200009)48:9/11<771::AID-PROP771>3.0.CO;2-E} {\bibfield
  {journal} {\bibinfo  {journal} {Fortschritte der Physik}\ }\textbf {\bibinfo
  {volume} {48}} (\bibinfo {year} {2000}),\
  10.1002/1521-3978(200009)48:9/11<771::AID-PROP771>3.0.CO;2-E}\BibitemShut
  {NoStop}%
\bibitem [{\citenamefont {Schoelkopf}\ and\ \citenamefont
  {Girvin}(2008)}]{Schoelkopf2008}%
  \BibitemOpen
  \bibfield  {author} {\bibinfo {author} {\bibfnamefont {R.~J.}\ \bibnamefont
  {Schoelkopf}}\ and\ \bibinfo {author} {\bibfnamefont {S.~M.}\ \bibnamefont
  {Girvin}},\ }\href {\doibase 10.1038/451664a} {\bibfield  {journal} {\bibinfo
   {journal} {Nature}\ }\textbf {\bibinfo {volume} {451}} (\bibinfo {year}
  {2008}),\ 10.1038/451664a}\BibitemShut {NoStop}%
\bibitem [{\citenamefont {Devoret}\ and\ \citenamefont
  {Schoelkopf}(2013)}]{Devoret2013}%
  \BibitemOpen
  \bibfield  {author} {\bibinfo {author} {\bibfnamefont {M.~H.}\ \bibnamefont
  {Devoret}}\ and\ \bibinfo {author} {\bibfnamefont {R.~J.}\ \bibnamefont
  {Schoelkopf}},\ }\href {\doibase 10.1126/science.1231930} {\bibfield
  {journal} {\bibinfo  {journal} {Science}\ }\textbf {\bibinfo {volume} {339}}
  (\bibinfo {year} {2013}),\ 10.1126/science.1231930}\BibitemShut {NoStop}%
\bibitem [{\citenamefont {Schuster}(2007)}]{Schuster2007}%
  \BibitemOpen
  \bibfield  {author} {\bibinfo {author} {\bibfnamefont {D.~I.}\ \bibnamefont
  {Schuster}},\ }\emph {\bibinfo {title} {Circuit Quantum Electrodynamics}},\
  \href@noop {} {\bibinfo {type} {{Ph.D.} thesis}},\ \bibinfo  {school} {Yale
  University} (\bibinfo {year} {2007})\BibitemShut {NoStop}%
\bibitem [{\citenamefont {Krantz}\ \emph {et~al.}(2019)\citenamefont {Krantz},
  \citenamefont {Kjaergaard}, \citenamefont {Yan}, \citenamefont {Orlando},
  \citenamefont {Gustavsson},\ and\ \citenamefont {Oliver}}]{Oliver}%
  \BibitemOpen
  \bibfield  {author} {\bibinfo {author} {\bibfnamefont {P.}~\bibnamefont
  {Krantz}}, \bibinfo {author} {\bibfnamefont {M.}~\bibnamefont {Kjaergaard}},
  \bibinfo {author} {\bibfnamefont {F.}~\bibnamefont {Yan}}, \bibinfo {author}
  {\bibfnamefont {T.~P.}\ \bibnamefont {Orlando}}, \bibinfo {author}
  {\bibfnamefont {S.}~\bibnamefont {Gustavsson}}, \ and\ \bibinfo {author}
  {\bibfnamefont {W.~D.}\ \bibnamefont {Oliver}},\ }\href {\doibase
  10.1063/1.5089550} {\bibfield  {journal} {\bibinfo  {journal} {Applied
  Physics Reviews}\ }\textbf {\bibinfo {volume} {6}},\ \bibinfo {pages}
  {021318} (\bibinfo {year} {2019})},\ \Eprint
  {http://arxiv.org/abs/https://doi.org/10.1063/1.5089550}
  {https://doi.org/10.1063/1.5089550} \BibitemShut {NoStop}%
\bibitem [{\citenamefont {Naghiloo}(2019)}]{Mahdi}%
  \BibitemOpen
  \bibfield  {author} {\bibinfo {author} {\bibfnamefont {M.}~\bibnamefont
  {Naghiloo}},\ }\href@noop {} {\enquote {\bibinfo {title} {Introduction to
  experimental quantum measurement with superconducting qubits},}\ } (\bibinfo
  {year} {2019}),\ \Eprint {http://arxiv.org/abs/1904.09291} {arXiv:1904.09291
  [quant-ph]} \BibitemShut {NoStop}%
\bibitem [{\citenamefont {Place}\ \emph {et~al.}(2021)\citenamefont {Place},
  \citenamefont {Rodgers}, \citenamefont {Mundada}, \citenamefont {Smitham},
  \citenamefont {Fitzpatrick}, \citenamefont {Leng}, \citenamefont {Premkumar},
  \citenamefont {Bryon}, \citenamefont {Vrajitoarea}, \citenamefont {Sussman},
  \citenamefont {Cheng}, \citenamefont {Madhavan}, \citenamefont {Babla},
  \citenamefont {Le}, \citenamefont {Gang}, \citenamefont {Jäck},
  \citenamefont {Gyenis}, \citenamefont {Yao}, \citenamefont {Cava},
  \citenamefont {de~Leon},\ and\ \citenamefont {Houck}}]{Place2021}%
  \BibitemOpen
  \bibfield  {author} {\bibinfo {author} {\bibfnamefont {A.~P.}\ \bibnamefont
  {Place}}, \bibinfo {author} {\bibfnamefont {L.~V.}\ \bibnamefont {Rodgers}},
  \bibinfo {author} {\bibfnamefont {P.}~\bibnamefont {Mundada}}, \bibinfo
  {author} {\bibfnamefont {B.~M.}\ \bibnamefont {Smitham}}, \bibinfo {author}
  {\bibfnamefont {M.}~\bibnamefont {Fitzpatrick}}, \bibinfo {author}
  {\bibfnamefont {Z.}~\bibnamefont {Leng}}, \bibinfo {author} {\bibfnamefont
  {A.}~\bibnamefont {Premkumar}}, \bibinfo {author} {\bibfnamefont
  {J.}~\bibnamefont {Bryon}}, \bibinfo {author} {\bibfnamefont
  {A.}~\bibnamefont {Vrajitoarea}}, \bibinfo {author} {\bibfnamefont
  {S.}~\bibnamefont {Sussman}}, \bibinfo {author} {\bibfnamefont
  {G.}~\bibnamefont {Cheng}}, \bibinfo {author} {\bibfnamefont
  {T.}~\bibnamefont {Madhavan}}, \bibinfo {author} {\bibfnamefont {H.~K.}\
  \bibnamefont {Babla}}, \bibinfo {author} {\bibfnamefont {X.~H.}\ \bibnamefont
  {Le}}, \bibinfo {author} {\bibfnamefont {Y.}~\bibnamefont {Gang}}, \bibinfo
  {author} {\bibfnamefont {B.}~\bibnamefont {Jäck}}, \bibinfo {author}
  {\bibfnamefont {A.}~\bibnamefont {Gyenis}}, \bibinfo {author} {\bibfnamefont
  {N.}~\bibnamefont {Yao}}, \bibinfo {author} {\bibfnamefont {R.~J.}\
  \bibnamefont {Cava}}, \bibinfo {author} {\bibfnamefont {N.~P.}\ \bibnamefont
  {de~Leon}}, \ and\ \bibinfo {author} {\bibfnamefont {A.~A.}\ \bibnamefont
  {Houck}},\ }\href {\doibase 10.1038/s41467-021-22030-5} {\bibfield  {journal}
  {\bibinfo  {journal} {Nature Communications}\ }\textbf {\bibinfo {volume}
  {12}} (\bibinfo {year} {2021}),\ 10.1038/s41467-021-22030-5}\BibitemShut
  {NoStop}%
\bibitem [{\citenamefont {Carroll}\ \emph {et~al.}(2021)\citenamefont
  {Carroll}, \citenamefont {Rosenblatt}, \citenamefont {Jurcevic},
  \citenamefont {Lauer},\ and\ \citenamefont {Kandala}}]{carroll2021dynamics}%
  \BibitemOpen
  \bibfield  {author} {\bibinfo {author} {\bibfnamefont {M.}~\bibnamefont
  {Carroll}}, \bibinfo {author} {\bibfnamefont {S.}~\bibnamefont {Rosenblatt}},
  \bibinfo {author} {\bibfnamefont {P.}~\bibnamefont {Jurcevic}}, \bibinfo
  {author} {\bibfnamefont {I.}~\bibnamefont {Lauer}}, \ and\ \bibinfo {author}
  {\bibfnamefont {A.}~\bibnamefont {Kandala}},\ }\href@noop {} {\enquote
  {\bibinfo {title} {Dynamics of superconducting qubit relaxation times},}\ }
  (\bibinfo {year} {2021}),\ \Eprint {http://arxiv.org/abs/2105.15201}
  {arXiv:2105.15201 [quant-ph]} \BibitemShut {NoStop}%
\bibitem [{zcu(2020)}]{zcu216}%
  \BibitemOpen
  \href@noop {} {\enquote {\bibinfo {title} {Xilinx and {A}vnet {ZCU}216
  evaluation board user guide},}\ }\bibinfo {howpublished}
  {\url{https://www.xilinx.com/support/documentation/boards_and_kits/zcu216/ug1390-zcu216-eval-bd.pdf}}
  (\bibinfo {year} {2020})\BibitemShut {NoStop}%
\bibitem [{ope(2021)}]{openqasm}%
  \BibitemOpen
  \href@noop {} {\enquote {\bibinfo {title} {{IBM} {Q}iskit {O}pen{QASM}
  library},}\ }\bibinfo {howpublished}
  {\url{https://github.com/Qiskit/openqasm}} (\bibinfo {year}
  {2021})\BibitemShut {NoStop}%
\end{thebibliography}%

\end{document}